\documentclass[prd,aps,preprintnumbers,floats,floatfix,superscriptaddress,preprintnumbers,showpacs,eqsecnum,nofootinbib,notitlepage,twocolumn]{revtex4} 

\usepackage[final]{graphicx}
\usepackage{dcolumn}
\usepackage{bm}
\usepackage{amsmath}	
\usepackage{amssymb}	
\usepackage{hyperref}
\usepackage{url}
\usepackage{float}
\newcommand{\beq}{\begin{equation}}
\newcommand{\eeq}{\end{equation}}
\usepackage{hyperref}
\newcommand{\nn}{\nonumber \\}
\newcommand{\bea}{\begin{eqnarray}}
\newcommand{\ena}{\end{eqnarray}}
\newcommand{\beann}{\begin{eqnarray*}}
\newcommand{\enann}{\end{eqnarray*}}
\newcommand{\gsim}{\, \mbox{\raisebox{-1.ex}
{$\stackrel{\textstyle>}{\textstyle\sim}$}}\,}

\begin{document}


\title{Gravitational Waves from Hierarchical Triple Systems 
with Kozai-Lidov Oscillation}

\author{Priti \sc{Gupta}} 
 \email{priti.gupta@tap.scphys.kyoto-u.ac.jp}
\affiliation{
Department of Physics, Kyoto University, Kyoto 606-8502, Japan
}
\author{Haruka \sc{Suzuki}} 
 \email{suzuki@heap.phys.waseda.ac.jp}
\affiliation{
Graduate School of Advanced Science and Engineering, Waseda University, Shinjuku, Tokyo 169-8555, Japan
}
\author{Hirotada \sc{Okawa}} 
 \email{h.okawa@aoni.waseda.jp}
\affiliation{
Waseda Institute for Advanced Study {\rm (WIAS)}, Waseda University, Shinjuku, Tokyo 169-8050, Japan
}
\author{Kei-ichi \sc{Maeda}} 
 \email{maeda@waseda.jp}
\affiliation{
Waseda Institute for Advanced Study {\rm (WIAS)}, Waseda University, Shinjuku, Tokyo 169-8050, Japan
}
\affiliation{
Department of Physics, Waseda University,
Shinjuku, Tokyo 169-8555, Japan
}

\date{\today}

\begin{abstract}
We study gravitational waves from a hierarchical three-body system up to
first-order post-Newtonian approximation. 
Under certain conditions, the existence of a nearby third
body can cause periodic exchange between eccentricity of an inner binary 
and relative inclination, known as Kozai-Lidov oscillations. 
We analyze features of the waveform from the inner 
binary system undergoing such oscillations. 
We find that variation caused due to the tertiary
companion can be observed in the gravitational waveforms and energy spectra,
which should be compared with those from isolated binaries and coplanar three-body system. 
The detections from future space-based interferometers will make possible the investigation of gravitational wave spectrum
in mHz range and may fetch signals by sources addressed.
\end{abstract}

\maketitle

\section
{\label{sec:1} INTRODUCTION}

The LIGO-VIRGO collaboration observed eleven gravitational wave signals from compact binary mergers during the first and second runs (O1 and O2)\citep{O1O2}. The third observation run (O3) by LIGO and VIRGO began in April 2019 and new detection alerts have started following within months of operation\citep{GWdata}. With current sensitivity design, ground detectors are focusing on the high-frequency range (10Hz to 1000Hz). 
Most expected sources lying in this range are compact binaries in the merger and ringdown phase.
For ground-based detectors, 
currently, we have five ground-based detectors with characteristic strain of order $10^{-22}$; advanced LIGO (aLIGO) detectors in Hanford and Livingston\citep{ligo}, advanced VIRGO (aVIRGO) in Italy\citep{virgo}, KAGRA in Japan\citep{kagra}, and GEO600 in Germany\citep{geo}.

Future space-based observatories like Laser Interferometer Space Antenna (LISA)  are expected to explore low frequency range from $10^{-4}$ Hz to $10^{-1}$ Hz with characteristic strain of order $10^{-21}$\citep{eLisa} whereas  DECi-hertz Interferometer Gravitational wave Observatory (DECIGO)
is aiming to fill the gap between LIGO and LISA with frequency band around $10^{-2}$ Hz to 10 Hz with characteristic strain down to $10^{-24}$\citep{decigo}. Several big projects have also been proposed in space; 
Big Bang Observer (BBO)\citep{bbo}, advanced LISA 
(aLISA)\citep{Larson2000}, and TianQin and TAIJI in China\citep{tq,taiji}.
With the launch of those space detectors,
we can aim for low-frequency sources. 
It will be possible to map a source all the way from inspiral to a merger phase by incoming gravitational waves. The observations so far, and the future GW detections may shed light on binary formation channels, enable precision tests of general relativity (GR) in the strong-field regime, and initiate new avenues in astrophysics.

In this new stage of the beginning of gravitational wave astronomy,
we have to list up all possible gravitational wave sources and 
analyze what we expect from those in observations.
So far, we have seen intensive analysis for binary systems composed of 
objects with various masses\citep[and references therein]{blanchet06,shibata11,faber12}.
However, we may expect more exotic sources in nature. 
One of the likely sources is a  three-body system, 
which we will study in this paper.
The environment near supermassive black holes (SMBHs) in galactic nuclei comprises of a large number of stars and compact objects. Triple systems could also emerge in these surroundings \cite[and references therein]{Heggie1975,Hut1993,Samsing2014,Riddle2015,Fabio2016,stephan2019} with the likelihood of the presence of hierarchical structure. The potential astrophysical sources of gravitational waves involving three-body system have seen increasing interest over time 
\citep{Wardell2002,Glutekin2006,Chiba07,Campanelli2008,Torigoe09,hoang18,Hoang2019}. All the detections so far (direct or indirect) came from binary inspirals and merger phase. It is good to contemplate how the waveform would look like if the binary system was accompanied by a tertiary companion in a hierarchical arrangement. 
The hierarchical triple body approximation has useful applications and can be applied to diverse settings, varying from planetary and stellar  mass scales to SMBHs. In this approximation, the energy of each orbit is separately conserved and therefore the two semi-major axes are constants. The Kozai-Lidov (KL) mechanism \citep{Kozai62, Lidov62},  which is one of the most important phenomenon in a hierarchical triple system, is particularly exciting. KL mechanism occurs when the inner orbit inclined with respect to the outer orbit.
The main feature of KL mechanism is the secular change (timescale much larger than the orbital periods and referred as KL timescale) of the eccentricity of the inner orbit and the relative inclination between inner and outer orbit. Both values oscillate inversely with respect to the  other, that is, when the eccentricity decreases, the inclination increases, and vice versa. The eccentricity of the orbit can reach extreme values leading to various  astrophysical phenomena.
For instance, if the eccentricity becomes large enough 
because of KL oscillations, the merger rate of BHs can be 
enhanced by large emission of gravitational waves \citep{Blaes02, Miller02,Wen2003,Liu17},  the tidal disruptions rate in a stellar cusp containing SMBH binary can be several orders  larger
in magnitude than expected in case of single SMBH \citep{Ivanov05, Chen09, Chen11, Wegg11, Li15,sari19}. Further, the formation of hot Jupiters or ultra-short-period planets is studied through the secular gravitational interactions between two planets in eccentric and inclined orbits \citep{Naoz12, Petrovich15, Anderson16,Oberst17}.

Recently, there has been extensive work focusing on gravitational wave detections from such systems. The probability of detection and frequency range from different sources has been discussed by many authors. The secular evolution of stellar BH (SBH) binaries near SMBHs is studied, focusing on the tidal perturbation by the SMBH \citep{Amaro-Seoane2010, Antonini2012,hoang18}. The analysis shows that merging binaries will enter the LIGO observational window while on orbits that are still very eccentric $(e \sim 0.5)$. The efficient GW analysis for such systems would, therefore, require the use of eccentric templates. The constraints are put on the importance of KL induced mergers for producing gravitational wave sources detectable by aLIGO, for triples with an inner pair of SBHs. 
Another interesting model with triple SMBHs is studied \citep{Antonini2016}. 
Numerical simulations of the dynamics of triples within galaxy cores exhibit phases of very high eccentricity (as high as $e \sim 0.99$) showing the likelihood of detection of GWs by LISA.

Some authors discuss variations in GWs in the presence of the third body due to relativistic beaming, Doppler, and gravitational redshift \citep{Meiron2017,Robson2018,Lisa2018,Lisa2019}. The acceleration imparted by the hierarchical companions leads to potentially observable changes in the waveforms that would reflect their gravitational interactions with the surrounding matter. The GW signal would be distinguished by a net phase change or even a time-dependent Doppler shift arising from the orbital motion. This could provide direct information about the black hole binary environments and otherwise invisible ambient mass. 

Recent work analyzes triple systems composed of the SMBH near the galactic center and a pair of BHs with different masses, from stellar to intermediate-mass BHs (IMBHs)\citep{Hoang2019,Loeb2019}. They identify frequency peaks and examine the detectability of GWs with LISA, $\mu$Ares, and DECIGO by integrating over the observation time within the lifetime of different GW detectors. Hoang et. al \citep{Hoang2019} show that the eccentricity oscillations of these binaries can be detected with LISA for BH binaries, with observation periods shorter than the mission lifetime, thereby disentangling this merger channel from others.

Indirect observation of GW from a triple system is also studied by analyzing the cumulative shift of periastron time of a binary pulsar undergoing KL oscillations \citep{Haruka2019}. 
 
In this paper, by integrating 
the equations of motion  directly, we analyze the features of the gravitational waves generated in a hierarchical triple system. We discuss wider range of parameters, i.e., 
a triple system of SBHs as well as a SBH binary 
around SMBH. We also analyze models composed of an IMBH as well as a SBH/SMBH.
We show the waveform and spectra as well as 
the observability of such sources by future space-based detectors. 
We also draw comparisons with the isolated binary systems. 

The paper is organized as follows: We review the characteristics of the hierarchical structure and KL-mechanism in \S\ref{sec:KL}. In section \S\ref{sec:Analytic} , we discuss the constraints on KL oscillations due to relativistic effects. In section \S\ref{sec:GRWs}, we recall the fundamentals of quadrupole formalism and discuss GWs from highly eccentric orbits. In section \S\ref{sec:classification}, 
 we explore parameter space for stable KL oscillations, classifying 
the cases by the mass ratios. We analyze seven models and 
show the properties of GWs for three typical models in \S\ref{sec:Results}.
We also discuss the observability of those models.
The paper is concluded in \S\ref{sec:Conc}. In Appendix \ref{waveform_IB_CTS}, we provide the waveforms from isolated binary systems and hierarchical triple coplanar systems for comparison.
We present the characteristic strains for all other models in Appendix \ref{Observability of other models}.

\section{Dynamics of Triple System}
\label{sec:Dynamics}
In this section, we sequentially explain a hierarchical triple system, Kozai-Lidov mechanism, and their stability and constraints.

\subsection{Hierarchical Triple System and Kozai-Lidov Mechanism}
\label{sec:KL}
\begin{figure}[h]
		\includegraphics[width=5cm]{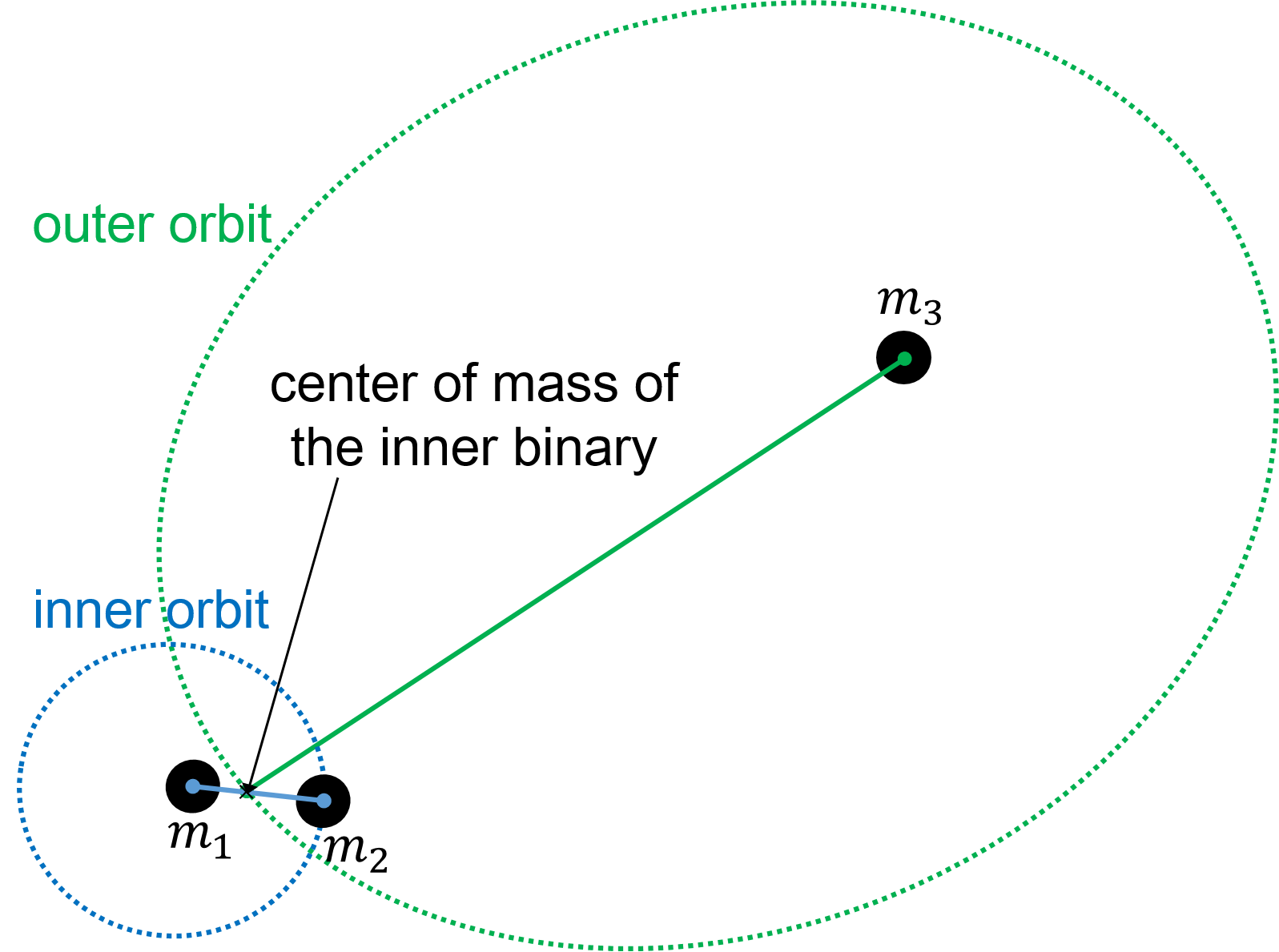}
\caption{The hierarchical triple system is constructed from inner and outer binaries.
					The inner binary consists of objects whose masses are $m_1$ and $m_2$, 
					and the outer one is the pair of  the inner binary  and 
                 the third body with mass $m_3$.}
		\label{fig:3body}
	\end{figure}

We can have several stable configurations and shapes when it comes to a three-body dynamical system. Broadly addressing, it can be concentric, coplanar orbits or inclined orbits in a hierarchical arrangement.
In this paper, we treat the so-called hierarchical triple system, whose schematic picture is given in Fig.~\ref{fig:3body}.
	
As depicted in the figure, we can decouple a system into the two-body inner binary orbit and the outer orbit with tertiary companion in motion around center of mass of inner binary, if  
the distance between the first and second bodies is much shorter than the distance to the third body. This hierarchy is supported by assuming that the gravitational effect of the third body is much smaller than the gravitational interaction between the first and second bodies.  

In a two-body problem in Newtonian dynamics, 
a bound system is given by an elliptic orbit, which
is described by six orbital elements; the semi-major axis $a$, the eccentricity $e$, the inclination $i$, the argument of periastron $\omega$, the longitude of ascending node $\Omega$, and the mean anomaly $M$.
Although these elements  are constant (except mean anomaly) in an isolated two-body system, in the hierarchical three-body system, the perturbations from the tertiary companion affect the binary motion and modify  the trajectory from that of the isolated one. Such a trajectory is not closed in general, but we can introduce an {\it osculating orbit} at each time, whose trajectory is approximated by the elliptical orbit with the above six orbital elements determined by the  instantaneous position and velocity \citep{Murray00}.
As for the outer orbit, we pursue the center of mass of the inner binary rotating around the tertiary companion (see Fig.~\ref{fig:3body}). It can also be described as another osculating orbit. Hence, we  introduce two osculating orbits, which are called as inner and outer orbits:
the masses of the inner binary are $m_1$ and $m_2$, while the tertiary companion has the mass $m_3$.  We use subscriptions `in' and `out' to show the elements of inner and outer orbits, respectively.

In the hierarchical three-body system with large relative inclination, 
a secular change of orbital elements may occur.
Under some conditions, there appears an oscillation between 
the relative inclination and the eccentricity of the orbit, which is called 
 Kozai-Lidov  (KL) oscillation\citep{Kozai62,Lidov62}.
This mechanism was independently discovered by Kozai and Lidov in 1962
\footnote{Although the framework and fundamental formulation had been already established by Von Zeipel in 1910 \citep{von1910,Ito2019}, we 
shall call it Kozai-Lidov mechanism because it is commonly used.}. 

The relative inclination $I$ is defined as the argument between the inner and outer orbital planes.
It is given by two orbital elements as
	\begin{eqnarray}
\cos I &=& \cos i_{\mathrm{in}} \cos i_{\mathrm{out}} 
				   + \sin i_{\mathrm{in}} \sin i_{\mathrm{out}} 
						\cos \left( \Omega_{\mathrm{in}} - \Omega_{\mathrm{out}} \right).
\nonumber
\\&&~
\label{eq:relative_i}
	\end{eqnarray}

If we assume $m_1\ll m_2, m_3$ and a circular outer orbit, we find that the oscillation occurs in secular time-scale under conservation of energy and angular momentum. 
In Newtonian quadrupole approximation method \citep{Shevchenko17}, it results in the secular exchange of  $e_\mathrm{in}$ and $I$ with the conserved value of $\Theta$, 
which is defined by 
	\begin{equation}
		\Theta = (1-e_\mathrm{in}^2) \cos^2 I \,.
		\label{eq:theta}
	\end{equation}
This approximation also gives the criterion of KL-oscillation for circular orbit as
	\begin{equation}
	0 \le |\cos I  |
			\ \le \sqrt{\frac{3}{5}} ,
		\label{eq:kozaiC}
	\end{equation}
which is equivalent to $39.2315^\circ \lesssim  I  
			\lesssim 140.7685^\circ $. 

This KL type oscillation will occur even if $m_1$ is not much smaller than $m_2, m_3$
\cite{Naoz12}.
When we include a relativistic effect, we also find the similar oscillation\cite{Naoz16,Wen2003}.
The effect of general relativity (GR) changes these KL- criterion 
as \citep{Migaszewski11}.
	\begin{equation}
 				0
		\le |\cos I|  \ 
		\le \sqrt{{3\over 5}\left(1-\frac{4}{9}
		\left\langle \mathcal{H}_\mathrm{GR} 
        \right\rangle\right) } ,
		\label{eq:KozaiC_PN}
	\end{equation}
where the GR correction term $\left\langle \mathcal{H}_{\rm GR} \right\rangle 	$
 is derived from the double-averaged post-Newtonian Hamiltonian of two-body relative motion \citep{Richardson88} as
      \begin{equation}       
		\left\langle \mathcal{H}_{\rm GR} \right\rangle 					
			= \frac{3Gm_{\rm in}^2a^3_{\rm out} \sqrt{1- e^2_{\rm out}}}{c^2 m_3a^2_ {\rm in} \sqrt{1- e^2_{\rm in}}}										\,,
		\label{eq:H_GR}
\end{equation}
where $m_{\rm in}=m_1+m_2$ is the total mass of the inner binary.
Since Eq. (\ref{eq:KozaiC_PN}) gives the stronger restriction, the GR effect may suppress the KL oscillation.
The KL-oscillation time-scale is also modified when the GR  effect is taken into account \cite{Naoz16}.

Note that it has been pointed out \cite{Will14a,Will14b} that 
it is imperative to take into account the
`cross terms' between the Newtonian perturbation and the post-Newtonian precession.
In our numerical evolution, we integrate the equations of motion (up to 1PN) directly. The  effects of the  cross terms are automatically included. 

\subsection{Stability and Constraints}
\label{sec:Analytic}
In this section, we discuss stability criteria for a hierarchical triple system.
We also give constraints for KL oscillation to occur, which are obtained from the conditions such that KL-oscillations are not suppressed by relativistic effects, i.e.,  the relativistic precession of the periastron of the inner binary and 
the Lense-Thirring (LT) precession effect due to the rapid rotation of heavy third body. 
\subsubsection{\textbf{\underline{Typical time scales}}}
Before discussing stability and constraints, 
we first present typical time scales of a hierarchical triple system.
The orbital periods of the inner and outer binaries 
are given by
\begin{eqnarray}
P_{\rm in} &=& 2\pi \sqrt{a_{\rm in}^3\over Gm_{\rm in}},
\label{Pin}
\\
P_{\rm out} &=& 2\pi \sqrt{a_{\rm out}^3\over Gm_{\rm out}},
\label{Pout}
\end{eqnarray}
where $m_{\rm out}=m_1+m_2+m_3$ is the total mass of a triple system.

The KL oscillation time scale is given by \citep{Antognini2015}
\begin{eqnarray}
	t_\mathrm{KL} \sim P_{\mathrm{in}}\frac{m_{\mathrm{in}}}{m_3} \left( \frac{a_{\mathrm{out}}}{a_\mathrm{in}} \right)^3
					 \left( 1-{e_\mathrm{out}}^2 \right)^\frac{3}{2}\,.
\label{tKL}
\end{eqnarray}
\subsubsection{\textbf{\underline{Stability of Hierarchical Arrangement}}}
For KL oscillations to occur, the first aim is to maintain the stability of a three-body system. A necessary (not sufficient) criteria is given by \citep{Aarseth01}, which predicts a minimum separation between inner and outer orbits for stability,
 beyond which a system may cause a chaotic instability, i.e., a system may become 
 either an unbounded system or a chaotic bounded system.
 Hence, we have one stability condition
 \footnote{There are also other relevant criteria for stability of a hierarchical system \citep{naoz13b,Petrovich15,hamers15}. Since the criteria are only necessary but not sufficient, the usefulness was questioned by Myllari et. al. \citep{Myllari18}. Their study takes into account the effect of inner binary eccentricity and relative inclination.},
\begin{equation}
		\frac{ a_\mathrm{out} }{ a_\mathrm{in} } >\frac{2.8}{ 1-e_\mathrm{out} } 
															\left[ \frac{m_{\rm out}}{m_\mathrm{in}}  
\frac{ 1+e_\mathrm{out} }{ (1-e_\mathrm{out})^\frac{1}{2} } \right]^\frac{2}{5}
\,.
\end{equation}
For simplicity, throughout this paper, we set outer orbit not to be so eccentric such that $e_{\rm out}\ll 1$. Thus the scaled expression is written as,
\begin{equation}
		\frac{ a_\mathrm{out} }{ {\rm AU} } > 2.8  
						\left(  \frac{m_{\rm out}}{M_\odot}\right)^{2\over 5}
						\left(\frac{m_\mathrm{in}}{M_\odot}\right)^{-{2\over 5}}               \, \frac{ a_\mathrm{in}}{ {\rm AU}} 
		\,
		\label{eq:max_stable}
	\end{equation}
This expression gives a lower bound to the separation between two orbits. Note that we have used this criterion just to set up initial data for the direct integration. Hence, the actual stability of our models is checked numerically.
\subsubsection{\textbf{\underline{Constraints from the relativistic effects}}}

\noindent
{\bf 3-1. Relativistic Precession of Periastron}\\[0.2em]
KL oscillations are suppressed or maximum eccentricity is restricted by the relativistic precession of the periastron of the inner binary. An analytical expression can be derived using double averaged approximation up to quadrupole order to estimate parameter space such that KL oscillations are not destroyed by precession\cite{Blaes02}.
The GR precession timescale of the periastron of the inner binary
can be estimated as,
\begin{eqnarray}
t_{\rm GR,in} \sim 
\bigg({\frac{a_{\rm in}}{r_{\rm g,in}}}\bigg)^{5\over 2} P_{\rm in}  \left( 1-{e_\mathrm{in}}^2 \right),
\end{eqnarray}
where $r_{g,{\rm in}}\equiv {Gm_{\rm in}/c^2}$ is the gravitational radius of the inner binary.
Comparing this with the KL time scale $t_{\rm KL}$, 
we find the constraint
\begin{equation}
\left({a_{\rm out}\over a_{\rm in}}\right)^3<
{3\over 4}
{a_{\rm in}\over r_{g,{\rm in}}}
{m_3\over m_{\rm in}}
\left({1-e_{\rm in}^2\over 1-e_{\rm out}^2}\right)^{3\over 2}
\,.
\end{equation}
For clarity, the scaled expression takes the form:
\begin{eqnarray}
&&\left({a_{\rm out}\over 
{\rm AU}}\right)^3<
10^8  \left({1-e_{\rm in}^2}\right)^{3\over 2}
\left({m_{\rm in}\over {M_\odot}} \right)^{-2} 
\left({a_{\rm in}\over {\rm AU}}\right)^4
\left(m_3 \over {M_\odot}\right)
\,,
\nonumber \\
&&
~~
\label{eq:GR_precession}
\end{eqnarray}
assuming $e_{\rm out}\ll 1$.\\


\noindent
{\bf 3-2 Lense-Thirring Precession Effect}\\[.2em]
Recent studies have shown changes in KL oscillations caused due to GR effects involving a SMBH \cite{Wang19,fang19,fang19a}.
If the third body is a rapidly rotating SMBH, then LT effect might become important changing the evolution of eccentricity excitation from the usual KL oscillation. It appears in the 1.5 PN order. 

For a rotating black hole, which is the third body in our setting,
the spin angular momentum is given as $S_3 = \chi_3 Gm_3^2/c$ where $\chi_3$ is the spin parameter, and outer orbital angular momentum is given as 
\begin{eqnarray}
L_{\rm out} = \mu_{\rm out} \left[G(1-e_{\rm out}^2)
a_{\rm out}M\right]^{1\over 2}
\,,
\end{eqnarray}
where the reduced mass of the outer orbit is defined by
\begin{eqnarray}
\mu_{\rm out} &=& {m_{\rm in} m_3\over m_{\rm out}}={(m_1+m_2) m_3\over m_1+m_2+m_3}
\,.
\end{eqnarray}

 The  time scale of the orbit-averaged precession 
 of $L_{\rm out}$ around $S_3$ is given as
\begin{equation}
t_{L_{\rm out}S_{3}} = \frac{GS_3(4+3m_{\rm in}/m_3)}{2c^2
(1-e^2_{\rm out})^{3\over 2} a_{\rm out}^3}
\end{equation}
The KL oscillations can be affected  when $t_{L_{\rm out}S_{3}}$ becomes comparable to the KL oscillation time scale $t_{\rm KL}$.

Hence, in order to observe the usual KL oscillation, we have the constraint such that 
$t_{L_{\rm out}S_{3}} > t_{\rm KL}$, for which
the scaled expression is given as:
\begin{equation}
\bigg({a_{\rm in}\over {\rm AU}}\bigg)^{3\over 2} >
10^{-12}
\bigg({m_3\over M_\odot} \bigg)  \bigg({m_{\rm in}\over M_\odot} \bigg)^{1\over 2}
\,,
\label{eq:LT_precession}
\end{equation}
assuming $\chi_3 \sim 1$.
If the ratio is comparable or greater than 1, then LT precession dominates resulting in chaotic evolution of eccentricity as shown in \cite{Wang19} and hence more dissipation of gravitational waves,
which may be interesting for further study.
However, in this paper our discussion is limited to 1PN effect and we focus more on the conventional KL oscillations in the inspiral phase.

\section{Gravitational Waves}
\label{sec:GRWs}
Under this section, we set the framework by recalling the famous quadrupole formula and definition of `near coordinate zone' (NCZ). We further discuss GWs from highly eccentric orbits, 
which are expected in KL oscillation.

\subsection{Quadrupole Formula}
\label{subsec:Quad}
In GR, the leading order contribution of the observed gravitational waves
is given by the quadrupole formula. It allows the evaluation of gravitational wave energy and waveforms emitted by a dynamical system.
In a wave zone, the gravitational waves denoted by a small deviation of the metric $h_{ij}$ are described by
\begin{equation}
\label{eq:QF}
  h_{ij}^{\rm TT}(t,\textbf{x}) = \frac{2G}{Dc^4} \ddot{Q}_{ij}^{\rm TT} 
  \,,
\end{equation}
where  TT refers to transverse-traceless gauge and a dot
denotes the time derivative.
 $D$ denotes the distance of the source from the observer.
The reduced quadrupole moment $Q_{ij}$ is defined by 
\begin{equation}
 Q_{ij} = M_{ij} - \frac{1}{3} \delta_{ij} M^k_{\,k}
 \,,
\end{equation}
where $i,j, \cdots$ run from 1 to 3 ($ x^1 = x , x^2=y,x^3=z $),
 and 
 the mass quadrupole moment $M_{ij}$ for $N$ masses is expressed as 
\begin{eqnarray}
M_{ij} = \sum_{A=1}^{N} m_A x_{A}^{i} x_{A}^{j}\,.
\end{eqnarray}
$m_A$ denotes the $A$-th mass at the location $x_A^i$ with origin at the center of mass of a system.

The power of GW emission is given by
\bea
{\cal P}={G\over 5c^5}\langle {\stackrel{\cdots}{Q}}_{ij}   
{\stackrel{\cdots}{Q}}\raisebox{.5em}{\scriptsize $ij$}\rangle
\,,
\ena
where the bracket denotes the Brill-Hartle averaging, which
is taken over several wave length \citep{Peters1963}.


The GW amplitude from a circular binary system consisting of masses $m_1$ and $m_2$ with orbital radius $a$ is approximately given by
\begin{eqnarray}
h \sim  {{4 G^2 m_1m_2}\over {c^4 a D}}
\,,
\end{eqnarray}
which is described by the scaled expression as
\begin{equation}
 h \sim 1.7\times 10^{-25} \bigg(\frac{m_1}{M_\odot}\bigg) \bigg(\frac{m_2}{M_\odot}\bigg)\bigg(\frac{{\rm AU}}{a_{\rm in}}\bigg) 
 \bigg(\frac{10{\rm kpc}}{D}\bigg)
\label{amplitude_binary}
\end{equation}
Another important quantity 
is energy spectra of the GWs.
For an isolated circular orbit, 
the energy radiated in gravitational waves is concentrated in the second harmonic such that 
\begin{equation}
 f_{\rm GW} = 2 f_{\rm in}\,,
\end{equation}
where $f_{\rm in}$ is the orbital frequency of the inner binary given by
\begin{equation}
f_{\rm in}={1\over P_{\rm in}}={1\over 2\pi}
\sqrt{Gm_{\rm in}\over a_{\rm in}^3}
\,.
\end{equation}
We find it in the energy spectrum, whose example is given in Appendix \ref{GW_isolated} and Fig. \ref{fig:spectra_binary}.

Large eccentricities induced in the inner binary during KL oscillation and variation of eccentricity with time produce more harmonics in gravitational waves, which make the energy spectra rich as compared to the spectra from an isolated circular binary as well as an enhancement of the wave amplitude 
as shown in the next subsection. 

In the derivation of the quadrupole formula,
a system must be contained in its NCZ , i.e., in order to use the quadrupole formula, a system should be inside a region (centered on the origin of the coordinates) of radius comparable to (or smaller than) the GW wavelength $\lambda$ (see some good example of what not to do given in \citep{Bonetti17}).
Therefore, we must keep a check that a hierarchical triple system is in its NCZ
 when the formalism is applied to the center of mass of the triple system for the model in consideration. There are two NCZs for a hierarchical triple system, whose sizes correspond to two wavelengths: one is that of an inner binary $\lambda_{\rm in} = a_{\rm in}c/v_{\rm in} $ and the other is that of an outer binary with  $\lambda_{\rm out} = a_{\rm out}c/v_{\rm out} $, where $v_{\rm in}$ and $v_{\rm out}$ are typical velocities of the inner binary and
the outer binary, respectively.
Since we focus on GWs from the inner binary, 
the NCZ condition is given by $a_{\rm out}<\lambda_{\rm in}$.
An inaccurate application may lead to apparent and 
unphysical behavior as shown in Fig. 18 of \citep{GP2011}.

\subsection{Gravitational waves from highly eccentric orbit}

In this subsection, we discuss
 the GWs from a highly eccentric binary system.
When the KL oscillation occurs, the inner binary system is 
in an inspiral phase with lower bound for the separation of the inner binary (see Eq. (\ref{eq:GR_precession})).
Since the separation is rather large, it is usually expected 
to be difficult to observe the GWs from the perspective both of 
the strain and the frequency. 
However, the eccentricity in the KL oscillation can become very large, 
which enhances the amplitudes of GWs and those frequencies.
As a result, we may have a chance to observe the GWs even in such 
a distant inspiral phase.

Gravitational Waveform from a binary system depends strongly on the eccentricity of the orbit.
In an eccentric orbit, the emission of gravitational waves is higher and  distributed over many harmonics. 
The GWs from an eccentric binary was studied by Peter and Mathews \citep{Peters1963}.
The total radiation power ${\cal P}$ averaged over one period is given by 
\bea
{\cal P}={32G^4m_1^2m_2^2m_{\rm in}\over c^5 a_{\rm in}^5(1-e_{\rm in}^2)^{7/2}}\left(1+{73\over 24}e_{\rm in}^2+{37\over 96}e_{\rm in}^4\right),
\ena
where $a_{\rm in}$ and $e_{\rm in}$ are the semi-major axis and the eccentricity of
 the inner binary, respectively.
 It shows that the radiation power increases 
 rapidly when the eccentricity becomes large.
 
The total radiation power ${\cal P}$ is the sum of the power radiated in the $n$-th harmonics
${\cal P}_n$ as
\bea
{\cal P}=\sum_{n=1}^\infty {\cal P}_n
\,.
\ena
The radiation power ${\cal P}_n$
 in the $n$-th harmonic is given by
\bea
{\cal P}_n={32G^4m_1^2m_2^2m_{\rm in}\over c^5 a_{\rm in}^5}g(n,e_{\rm in})
\ena
where
 \bea
 g(n,e)&=&{n^4\over 32}\Big{[}
 \Big{(}J_{n-2}(ne)-2eJ_{n-1}(ne)
 \nn
 &+&{2\over n}J_n(ne)+2eJ_{n+1}(ne)-J_{n+2}(ne)\Big{)}^2
 \nn
 &+&(1-e^2)\Big{(}J_{n-2}(ne)-2J_n(ne)+J_{n+2}(ne)\Big{)}^2 
 \nn
 &+&{4\over 3n^2}\left(J_n(ne)\right)^2\Big{]}
 \,,
 \ena
 where $J_n(x)$ is the Bessel function of the first kind. 
  
The spectrum ${\cal P}_n$ is peaked at the GW frequency $f_{\rm peak}$,
\begin{equation}
f_{\rm peak} = n_{\rm peak} f_{\rm in},
\end{equation}
where the magnification factor $n_{\rm peak}$ is well approximated by 
\begin{equation}
n_{\rm peak} = \frac{2 (1+ e)^{1.1954}}{(1- e^2)^{1.5}}
\,,
\end{equation}
for the range of $10^{-6}<1-e^2<1$ \citep{Wen2003}.
For $e \ll 1$, $n_{\rm peak}$ is just 2, while considering high eccentricity, say $e \sim 0.98$ , we find $n_{\rm peak} \sim 574$. 
As the eccentricity increases, change in $f_{\rm peak}$ becomes significant and may lie in the detectable frequency band $10^{-4}$ Hz $\sim 10$ kHz.
It is worthwhile evaluating the  peak frequency as
\begin{equation}
f_{\rm peak} \sim  3\times 10^{-8} n_{\rm peak} \bigg(\frac{m_{\rm in}}{M_\odot}\bigg)^{1\over2}\bigg(\frac{a_{\rm in}}{\rm AU}\bigg)^{-{3\over2}}~{\rm Hz}.
\end{equation}
Note that 
although the $n$-th harmonics power ${\cal P}_n$ 
has a peak at 
$f=f_{\rm peak}$, the spectrum is rather broad.
As a result, the observable range of the frequency 
can be wide as shown in \citep{Lisa2019}.

In observation, another important quantity is the wave amplitude.
The GW amplitude $h_n$ of the $n$-th harmonic is related to
${\cal P}_n$ by 
 \begin{eqnarray}
 \langle \dot{h}_n^2 \rangle  =(2\pi f_n)^2  \langle {h}_n^2\rangle \approx 
 \frac{4G}{c^3 D^2}{{\cal P}_n(a,e)}
 \,,
 \end{eqnarray}
where the bracket denotes the time average over a few orbital period, and $f_n$ is the frequency of the $n$-th harmonic wave.
\begin{widetext}
When we evaluate the amplitude at the peak frequency $f_n=f_{\rm peak}$,
we find 
\bea
\sqrt{\langle {h}_n^2\rangle}\approx 5\times 10^{-25}\,
{g^{1/2}(n_{\rm peak},e)\over n_{\rm peak}} 
\left({m_1\over M_\odot}\right)\left({m_2\over M_\odot}\right)\left({a_{\rm in}\over{\rm AU}}\right)^{-1}
\left({D\over 10{\rm kpc}}\right)^{-1}
\,.
\ena
\end{widetext}
Suppose that $m_1=m_2=10 M_\odot$, $a_{\rm in}=0.01$ AU, and $D=10{\rm kpc}$.
When the eccentricity gets large enough via KL oscillations, 
we obtain
\bea
\sqrt{\langle {h}_n^2\rangle} \sim 1.9\times 10^{-22}\,,
\ena
for $e_{\rm in}=0.98$.
It may be observable.

To answer the question whether it is really observable  by 
future space-based detectors, 
we have to perform a numerical simulation and 
evaluate the dimensionless characteristic strain,
which is defined as \cite{Cornish2018},
\begin{equation}
h_c(f) = 2 f 
\left(|\widetilde{h}_+(f)|^2+|\widetilde{h}_{\times}(f)|^2
\right)^{1\over 2}.
\end{equation}
To calculate this quantity, we first find gravitational waveforms $h_+(t)$ and $h_{\times}(t)$ and transform them into the Fourier components
$\widetilde{h}_+(f)$ and $\widetilde{h}_{\times}(f)$.
In our analysis, 
we apply Fast Fourier Transform (FFT) to perform the Fourier transformation, 
and  evaluate the characteristic strain.

The total GW energy 
\bea
E=\int dt\, {\cal P}
\ena
consists of many harmonics, which can be described as
\bea
E=\int_0^\infty df\, {dE\over df}
\,.
\ena
The energy spectrum ${dE\over df}$ 
is obtained as
\begin{equation}
{dE\over df}(f) = \frac{2 \pi^2 c^3 f^2 D^2}{5G} \left(|\widetilde{h}_+(f)|^2+|\widetilde{h}_{\times}(f)|^2
\right)
\,.
\end{equation}

\vskip .5cm
\section{Classification and Constraints on Parameters}
\label{sec:classification}

Before we perform our analysis, we classify the possible models with the KL oscillations by the masses of a triple system.
We focus on stable hierarchical triple systems and list the models in three different mass ranges. 
This general discussion helps in visualizing several realistic situations varying on the scale of mass.

We have three possible cases:\\
{\bf Case (A):} $m_{\rm in} \sim m_3$, in which three bodies have comparable masses, or one of the inner binary system can have a smaller mass. \\
{\bf Case (B):} $ m_{\rm in} \ll m_3$, which mimics the state of a less massive inner binary around a more massive BH, e.g., a SMBH in the galactic nuclei.
The inner binary can be a stellar mass binary system such as a binary BH,  a binary neutron star or a neutron star-black hole binary. It can also be a binary BH consisting of an IMBH.\\
{\bf Case (C):} $ m_{\rm in} \gg m_3$, which  is the reverse of Case (B), hence, instead of SMBH, we may have a planetary object or an asteroid as a tertiary companion. For instance, the inner binary consists of IMBHs, while the third companion is stellar mass size.

In Case (C), although the KL oscillation is possible for appropriate parameters, the oscillation occurs between the outer longitude of the ascending node and the relative inclination.
The eccentricity of the inner binary may not be affected so much. As a result, the gravitational waves from the inner binary system, which are much more important than those from the outer orbit,  has almost no effect due to KL oscillation.
We then do not analyze Case (C) in this paper.
There is another reason we do not discuss Case (C), i.e. 
KL timescale is much longer and the cycle may not be captured 
in one observation run.\\ 
 A recent work discusses the presence of SMBH binary in our galactic centre including the possibility of GW detection\citep{Naoz2020}.

Now, we show the parameter range in which the KL oscillation appears and is stable.

\subsection{Case (A) : $m_{\rm in}\sim m_3$}

\begin{figure}[h]
  \includegraphics[width=8.5cm]{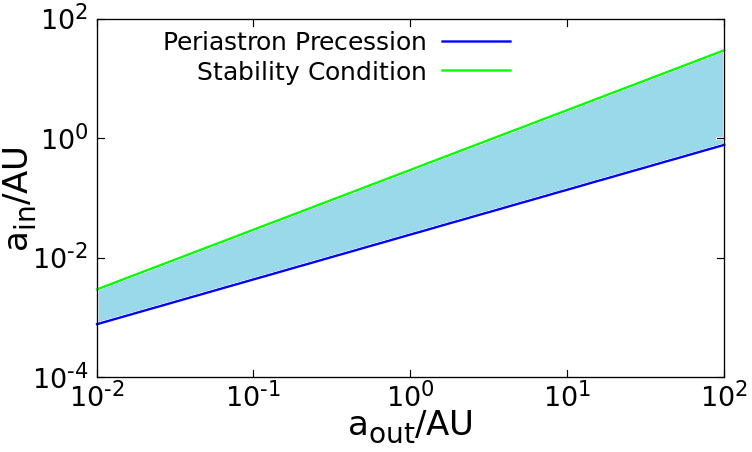}
\caption{\small
The parameter space $(a_{\rm out},a{\rm in})$ for Case (A). 
We set $m_1=m_2=m_3 = 10 M_\odot$.  
The relativistic effect (the precession of periastron) is shown by the blue line, while the stability condition (chaotic boundary) is given by green line.  The light-blue shaded region depicts the allowed parameter space for stable KL oscillations.}
\label{fig:parameter_ain_case_a}
\end{figure}

As discussed in \S \ref{sec:Analytic}, we have stability conditions and constraints in order to find the KL oscillations. 
In this case, the relation between the typical time scales are given as
\bea
{P_{\rm in}\over t_{\rm KL}} \ll {P_{\rm out}\over t_{\rm KL}}\sim {P_{\rm in}\over P_{\rm out}}
\ll 1
\ena
We present some examples in Fig. 
 \ref{fig:parameter_ain_case_a}, in which
we show the allowed region in the parameter space 
$(a_{\rm in}, a_{\rm out})$ for fixing the the masses 
of a triple system 
as $m_1=m_2=m_3=10 M_\odot$.
In \S \ref{sec:Analytic}, we show the condition with which the orbit does not become chaotic. This gives the maximum value of $a_{\rm in}$, which is given by green line [Eq.  (\ref{eq:max_stable})]. 
The other two constraints come from the relativistic effects. One is the precession of the periastron, whose condition is 
shown by the blue line [Eq. (\ref{eq:GR_precession})].  Beyond this line, such relativistic effect will suppress the KL oscillation. 
The other is the LT precession effect, whose 
critical line is out of the present parameter range.
Taking into account the above conditions,  in Fig. \ref{fig:parameter_ain_case_a}, we present the possible range in the parameter space of $(a_{\rm in}, a_{\rm out})$ for the stable KL oscillation, which is shown by the light-blue shaded region. 
For the present parameter choice, the constraints are determined by the relativistic precession effect of the periastron and chaotic boundary. Similar figure has also been presented by Naoz et. al. discussing the parameter range and different constraints \citep{naoz13b}. 

\subsection{Case (B) : $m_{\rm in}\ll m_3$}
In this case, the relation between the typical time scales are given as
\bea
 \left({P_{\rm in}\over t_{\rm KL}}\right)^{1/2} \sim {P_{\rm out}\over t_{\rm KL}}\sim {P_{\rm in}\over P_{\rm out}}
 \,.
\ena

\begin{figure}[h]
  \includegraphics[width=8.5cm]{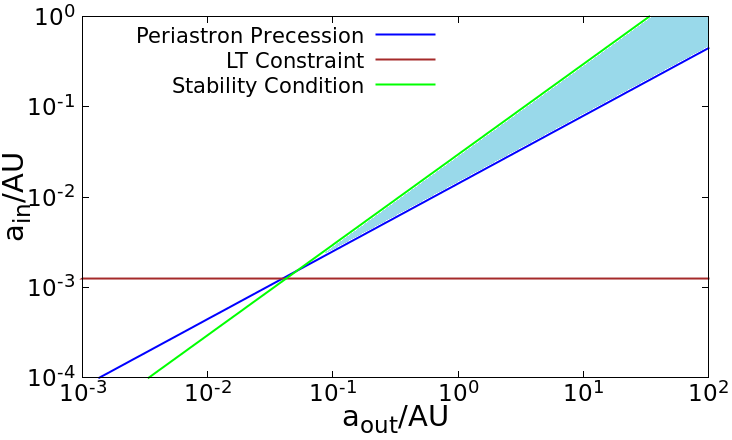}
\caption{\small
 We plot the allowed region (light-blue shaded) 
in the parameter space of $(a_{\rm in}, a_{\rm out})$ for the stable KL oscillation by fixing  $m_1=10^3 M_\odot , m_2= 10^3 M_\odot$ and 
$m_3 =  10^6 M_\odot$.
The blue and brown lines correspond to the critical curves of relativistic effects (precession of periastron and LT effect),
 beyond which the KL oscillation
 might be suppressed. The stability of the KL oscillation gives 
  one condition (chaotic boundary) shown 
  by green  line.}
\label{fig:parameter_ain_case_b}
\end{figure}

In Fig. \ref{fig:parameter_ain_case_b}, 
we present the possible range in the parameter space of $(a_{\rm in}, a_{\rm out})$ for 
fixing the  masses of  a triple system 
as $m_1=m_2=10^3 M_\odot$,  and $m_3=10^6 M_\odot$. 
The constraints are determined only by the relativistic precession effect of the periastron and 
chaotic boundary
just as Fig. \ref{fig:parameter_ain_case_a} in Case (A).
We find that there exists the minimum values of 
$a_{\rm in}$ and $a_{\rm out}$, below which 
no stable KL oscillation is possible.

\begin{figure}[h]
  \includegraphics[width=8.5cm]{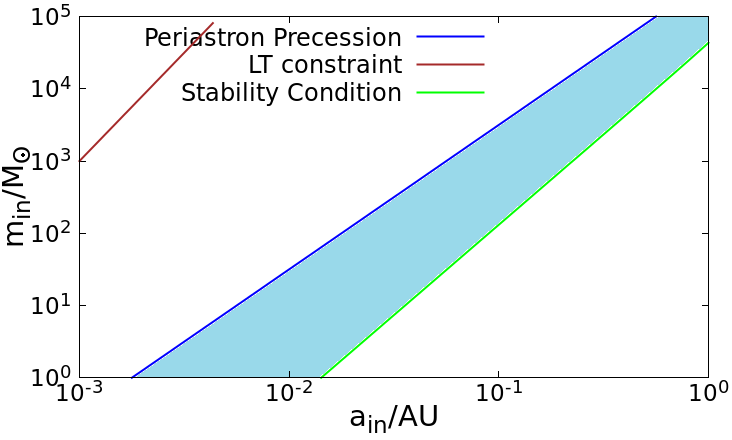}
\caption{\small 
 We plot the allowed region (light-blue shaded) 
in the parameter space of $(a_{\rm in}, m_{\rm in})$ for the stable KL oscillation by fixing  $m_3 = 10^6 M_\odot$ and $a_{\rm out} = 10 {\rm AU}$. }
\label{fig:parameter_case_b}
\end{figure}

Since the range of mass parameter in Case  (B) is very wide,
in Fig. \ref{fig:parameter_case_b}, we also present the possible range in the parameter space of $(a_{\rm in}, m_{\rm in})$ for the stable KL oscillation, which is shown by the light-blue shaded region. We fix the  mass of tertiary component and the semi-major axis of the outer orbit as $m_3=10^6 M_\odot$ and $a_{\rm out}=10 {\rm AU}$. 
For the present parameter choice, 
the constraints are determined by the relativistic precession effect of the periastron and chaotic boundary condition
just as Case (A).

\begin{figure}[h]
  \includegraphics[width=8.5cm]{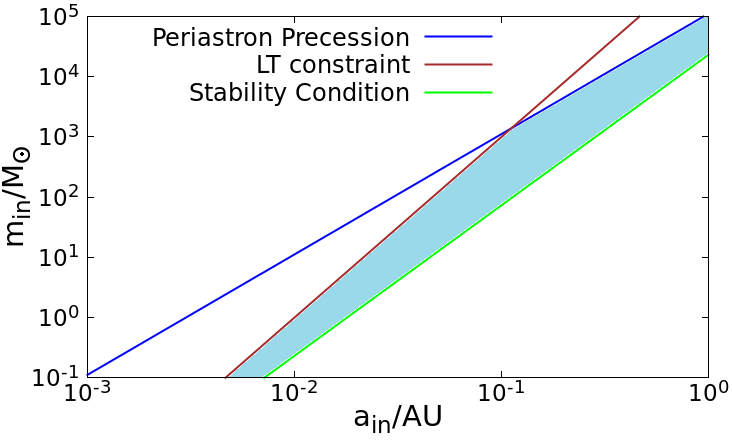}
\caption{\small 
The same figure as Fig. \ref{fig:parameter_case_b}
with the different parameters. 
We assume  $m_3 = 10^9 M_\odot$ and $a_{\rm out} = 200 {\rm AU}$. 
The LT critical curve [Eq. (\ref{eq:LT_precession})] cross the 
light-blue shaded region for stable KL oscillations. 
 }
\label{fig:parameter_case_b_10^9}
\end{figure}

However, note that the LT precession may become significant 
in Case (B) when the third body is much heavier.
See Fig. \ref{fig:parameter_case_b_10^9}, in which we
set $m_3=10^9 M_\odot$.
In this case, a hierarchical triple system will evolve 
in the left direction horizontally as 
$a_{\rm in}$ decreases because of the emission of
GWs. If $m_{\rm in}\lesssim 10^3 M_\odot$, the evolution
curve will hit on the LT critical line before the 
relativistic periastron shift becomes important.
Hence the system will evolve into a chaotic KL oscillation
phase, which may be very interesting to study, but it is
out of our present analysis because the LT precession appears 
in 1.5 PN order.

\section{Waveform, Spectra and Observability of GWs with KL Oscillations}
\label{sec:Results}
\subsection{Models}
To clarify the properties of the GWs from an inner binary in a hierarchical triple system, we have performed numerical 
analysis for seven models listed in Table \ref{models}.
 We assume that there exist three types of black holes as GW sources; 
stellar BH (SBH), intermediate-mass BH (IMBH) and supermassive BH (SMBH).
As typical masses of SBH, IMBH and SMBH, we choose $10 M_\odot, 10^3 M_\odot$ and $10^6 M_\odot$, respectively.
We have not included the models with two or more SMBHs.
Although those models can be important  in astrophysics or cosmology, their KL oscillation timescale is too long from the observational view point
(see Table \ref{model_and_observability} in Appendix
\ref{Observability of other models}). 
\begin{table}[H]
\begin{center}
  \begin{tabular}{|c||c|c|c||c||c|}
\hline 
&&&&&
\\[-.5em]
Inner Binary&$m_1[M_\odot]$&$m_2[M_\odot]$&$m_3[M_\odot]$&Case&Model
\\[.6em]
\hline 
&$10 $&$10  $&$10  $&A&IA1
\\[.1em]
\cline{2-6}
SBH-SBH&$10  $&$10  $&$10^3  $&B&IB3
\\[.1em]
\cline{2-6}
&$10  $&$10  $&$10^6  $&B&IB6
\\[.1em]
\hline 
\hline 
\raisebox{-.8em}{SBH-IMBH}&$10  $&$10^3  $&$10^3  $&A&IIA3
\\[-.5em]
\cline{2-6}
&$10  $&$10^3  $&$10^6  $&B&IIB6
\\[.1em]
\hline
\raisebox{-.8em}{IMBH-IMBH}&$10^3  $&$10^3  $&$10^3  $&A&IIIA3
\\[-.5em]
\cline{2-6}
&$10^3  $&$10^3  $&$10^6  $&B&IIIB6
\\[.1em]
\hline 
\hline 
  \end{tabular}
    \caption{SBH, IMBH and SMBH denote a stellar BH, an intermediate-mass BH, and a super massive BH, respectively.
    The last numbers in the model names describe the exponents of those masses.}
\label{models}
\end{center}
\end{table}

Since the basic features of the GWs are similar for all models, we shall discuss three representative models; 
Model IA1, IB3 and IB6, in which the inner binary
consists of 10 $M_\odot$ BHs.
We choose the semi-major axis $a_{\rm in}$ and 
$a_{\rm out}$ in the stable regions for the KL oscillation
(for example, the light-blue shaded region in Fig. \ref{fig:parameter_ain_case_a}).

In addition, to confer the results of our analysis, we pick Case (A) to contrast the key features in waveforms from an isolated binary system  (Models I$_{\rm C}$ and I$_{\rm E}$) and in
waveforms from a coplanar hierarchical triple system
(Models IA1$_{\rm CC}$ and IA1$_{\rm CE}$). The parameters are summarized in Table ~\ref{tab:ini}.

As we focus on the inspiral phase of the inner binary, there are special features that can be seen in the waveform.
 The observer's location $(\iota)$ is defined as the angle between the detector  and the ``axis" of the reference plane (i.e., the ``normal" to the initial outer orbital plane).

\begin{widetext}
~
\begin{table}[h]
\begin{tabular}{c|c|c|c||ccc|ccccc|ccc}	
\hline
 System& Case& Model  & Orbital motion   &$m_1$&$m_2$&$m_3$&   $a_{\rm in}$[{\rm AU}] &$a_{\rm out}$[{\rm AU}]  &  $e_{\rm in}$  &     $e_{\rm out}$  &$I$[deg] &$P_{\rm in}$[days]&$P_{\rm out}$[days]&$t_{\rm KL}$[days] 
\\
\hline\hline
&&IA1&   KL oscillation & 10&10&10&    0.01  &   0.1  &  0 & 0 & 90 &0.082& 2.10&$\sim$ 176\\
&(A) &IA1$_3$&   KL oscillation & 30&30&30&    0.01  &   0.1  &  0 & 0 & 90 & 0.047&
1.22&$\sim$ 123\\
\cline{3-15}
hierarchical &  &IA1$_{\rm CC}$& coplanar circular  & 10&10&10&       0.01  &   0.1  &  0 & 0 & 0& 0.082& 2.10&- \\
 ~~triplet  &  &IA1$_{\rm CE}$& coplanar eccentric  &  10&10&10&      0.01  &   0.1  &  0.9 & 0 & 0&0.082&2.10&-
\\
\cline{2-15}
 &(B)& IB3& KL oscillation  &  $10$&$10$&$10^3$&      0.01  &   0.5  &  0 & 0 & 90 & 0.082& 4.04&$\sim$ 255\\
 & &IB6& KL oscillation  &  $10$&$10$&$10^6$&    0.01  &   5 &  0 & 0 & 90 & 0.082&
 4.08&$\sim$ 96\\
\hline
isolated& & I$_{\rm C}$   & 
circular   & 10&10&-&        0.01  &   -  &  0 & - & - &0.082&-&-\\
 ~~binary &&I$_{\rm E}$  &  eccentric    & 10&10&-&       0.01  &   -  &  0.9 & - & - & 0.082&-&-\\
\hline
\end{tabular}
\caption{\small Masses of three-body hierarchical systems and of two body isolated binary, and the initial orbital elements $a$, $e$, and $i$ 
are the semi-major axis, eccentricity, and inclination respectively. Subscripts ``in" and ``out" correspond to the inner and outer orbits,  respectively. We also 
give the typical time scales; $P_{\rm in}$, $P_{\rm out}$ and $t_{\rm KL}$.
The eccentricity for KL binary 
increases maximally to $e_{\rm in, max}\sim 0.987$
and $e_{\rm in, max}\sim 0.964$ by the KL oscillation for models IA1 and IA1$_3$, while $e_{\rm in, max}\sim 0.99$ and $e_{\rm in, max}\sim 0.99$
for Models IB3 and IB6, respectively.}
\label{tab:ini}
\end{table}

\subsection{Method of Orbit Evolution}
The general relativistic equations of motion known as the following 
Einstein-Infeld-Hoffmann (EIH) equations \cite{EIH38}
describes the dynamics of a system of point-like masses due to their mutual gravitational interactions, including the first order post-Newtonian GR effects: 

		\begin{eqnarray}
&&	\frac{ \mathrm{d} \bm{v}_{k}}{\mathrm{d} t}
			=-G\sum_{n\neq k} m_{n}\frac{\bm{x}_{k} - \bm{x}_{n}}{|\bm{x}_{k} - \bm{x}_{n}|^{3}}
				 \Big[ 
						1-4 \frac{G}{c^2}\sum_{n'\neq k} \frac{m_{n'}}{|\bm{x}_{k} - \bm{x}_{n'}|}
                  		-\frac{G}{c^2}\sum_{n'\neq n} \frac{m_{n'}}{|\bm{x}_{n} - \bm{x}_{n'}|} 
						\left \{
                             		   1-\frac{(\bm{x}_{k} - \bm{x}_{n}) \cdot (\bm{x}_{n} - \bm{x}_{n'})}
                                     {2|\bm{x}_{n} - \bm{x}_{n'}|^{2}} 
						\right \} 
		\nonumber	 \\
&&						+\left( \frac{|\bm{v}_{k}|}{c} \right)^{2} + 2\left( \frac{|\bm{v}_{n}|}{c} \right)^{2} 
						-4 \frac{\bm{v}_{k} \cdot \bm{v}_{n}}{c^2}
                  		-\frac{3}{2} 
                  		\left\{ 
								   \frac{(\bm{x}_{k} - \bm{x}_{n})}{|\bm{x}_{k} - \bm{x}_{n}|} \cdot \frac{\bm{v}_{n}}{c} 
						\right\}^{2}   
                 \Big]  
                 -\frac{G}{c^2} \sum_{n \neq k} \frac{m_{n}(\bm{v}_{k}-\bm{v}_{n})}{|\bm{x}_{k} - \bm{x}_{n}|^{3}}
                		(\bm{x}_{k} - \bm{x}_{n}) \cdot (3\bm{v}_{n}-4\bm{v}_{k}) 
			\nonumber \\	\hspace*{-15cm}
&&			-\frac{7}{2} \frac{G^{2}}{c^2} \sum_{n \neq k}\frac{m_{n}}{|\bm{x}_{k} - \bm{x}_{n}|}
                		\sum_{n'\neq n} \frac{m_{n'} (\bm{x}_{n} - \bm{x}_{n'})} {|\bm{x}_{n} - \bm{x}_{n'}|^{3}},
		\label{eq:EIH}
\end{eqnarray}
\end{widetext}
 where  $m_k$, $\bm{v}_k$, $\bm{x}_k$ ($k=1,2$ and $3$)
are the mass, velocity and position of the $k$-th component of the system, $G$ is the gravitational constant, and $c$ is the speed of light.\footnote{ This equation could also be derived from the Lagrangian given by Lorentz and Droste \cite{LD17}.}

We employ the EIH equations  in order to solve the three-body system.
Eq.~(\ref{eq:EIH}) has been numerically integrated by using 6-th order implicit Runge-Kutta method, whose coefficients are obtained from \citet{Butcher64}.

On integrating, we obtain the numerical data of positions and velocities of the triple system. 
 In order to set up initial conditions, 
we convert initial orbital elements of inner and outer orbits
 into the variables $\bm{x}_{k}$ and $\bm{v}_{k}$ in Cartesian coordinates,
 with its origin in the center of mass of the whole system \citep{Murray00}.
 The initial outer orbital plane is considered to be the reference frame.
We integrate the EIH equations (\ref{eq:EIH}) numerically and 
then evaluate the osculating orbital elements at each step from the 
numerical data of positions and velocities of the triple system (see e.g. \citet{Murray00}). 
Since the inner orbit is not exactly an ellipse, 
the obtained osculating elements are oscillating with small amplitudes in the cycle of inner orbit. 
Hence, we take an average of the osculating elements for each cycle of inner orbit and then obtain the averaged semi-major axes $\bar{a}_\mathrm{in}$, $\bar{a}_\mathrm{out}$ and eccentricities $\bar{e}_\mathrm{in}$,  $\bar{e}_\mathrm{out}$, 
which may give the effective values of the orbital elements. 
Those elements will evolve secularly in time because of the effect of the tertiary body.

\subsection{Numerical Results}
We first focus on the numerical results for Model IA1
in next subsections {\it 1, 2} and {\it 3}, and 
then discuss about Model IB3 in subsection {\it 4}.

\subsubsection{\textbf{\underline{Evolution of the eccentricity}}}
\label{eccentricity}
We first show the result of the evolution of eccentricity and relative inclination of the inner binary for Model IA1, in which the masses
of a triplet are $m_1=m_2=m_3=10 M_\odot$ and initial semi-major axes are $a_{\rm in}=0.01 {\rm AU}$ and $a_{\rm out}=0.1 {\rm AU}$.
$P_{\rm in}$ (= 0.082 days) and $P_{\rm out}$(= 2.10 days) are given by the formulas (\ref{Pin}) and (\ref{Pout}), while 
$t_{\rm KL}$ ($\sim 176$ days)
is evaluated by the time interval between first and second peaks of 
the inner eccentricity.

\begin{figure}[h]
    \centering
    \includegraphics[width=6cm]{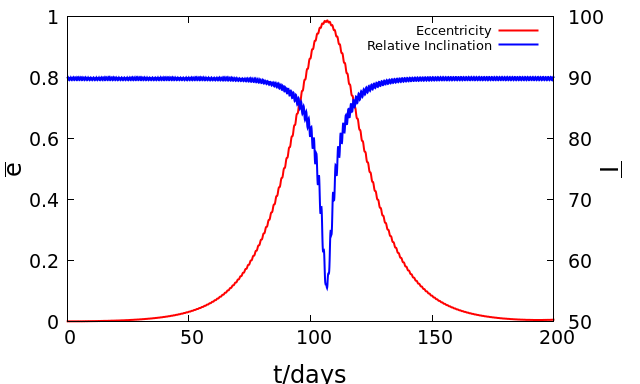}
    \caption{\small The time evolution of the osculating orbital elements
    (the averaged eccentricity of the inner binary 
    $\bar{e}_{\rm in}$, and averaged relative inclination $\bar{I}$)
    in Model IA1 ($m_1=m_2=m_3=10M_\odot$, $a_{\rm in}=0.01 {\rm AU}$, and $a_{\rm out}=0.1 {\rm AU}$)  for one KL oscillation cycle. The red and blue lines show the evolution of $\bar{e}_{\rm in}$ and $\bar{I}$, respectively. 
    When the inclination drops to $I_{\rm min}\approx 55^\circ$,
    the eccentricity reaches the maximum value $e_{\rm max}\approx 0.987$. The small oscillations are caused by the outer companion’s motion.}
    \label{fig:KL}
\end{figure}

The initial eccentricities are $e_{\rm in}=e_{\rm out}=0$, but 
the relative inclination is chosen to be $I=90^\circ$.
We show the time evolution of the eccentricity $e_{\rm in}$ and the relative inclination $I$ for one KL oscillation cycle 
 in  Fig ~\ref{fig:KL}. 
We have run our code for several values of  inclination varying from 40$^\circ$ to 90$^\circ$, for which we obtain stable KL oscillations.
For $I=90^\circ$, we find the maximum value of the eccentricity as $e_{\rm in, max}\approx 0.987$, while $e_{\rm in, max}\approx 0.02$ for $I=40^\circ$.
 For lower values for relative inclination, we find lower maximum eccentricities as expected from Eq. (\ref{eq:theta}).
In this paper, we show the result for $I=90^\circ$,
which gives the largest strain of the GWs, but 
we find that even for 
the case of $I=50^\circ$, in which the maximum 
eccentricity is about 0.42, the strain of the GWs 
can reach the observable sensitivity of LISA (see 
the yellow curve ($e_{\rm in}=0.361$) in Fig. \ref{fig:hfplot_A}).

\subsubsection{\textbf{\underline{GW Waveform}}}
\label{GW Waveform}
Next we show the waveform of the GWs for Model IA1
($m_1=m_2=m_3=10M_\odot$, $a_{\rm in}=0.01 {\rm AU}$, and $a_{\rm out}=0.1 {\rm AU}$).
\begin{figure}[h]
  \centering
  \includegraphics[width=0.70\linewidth]{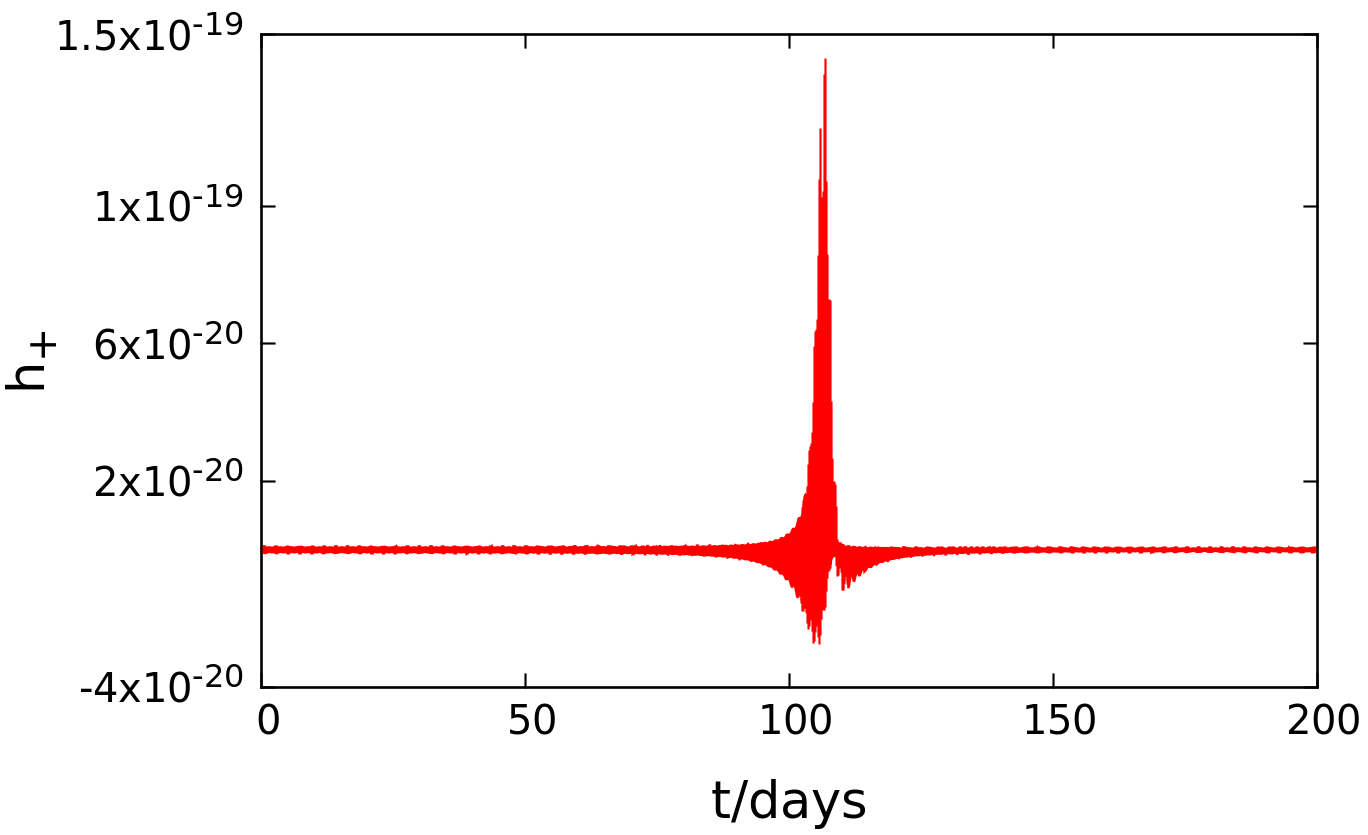}
  \\
  \includegraphics[width=0.70\linewidth]{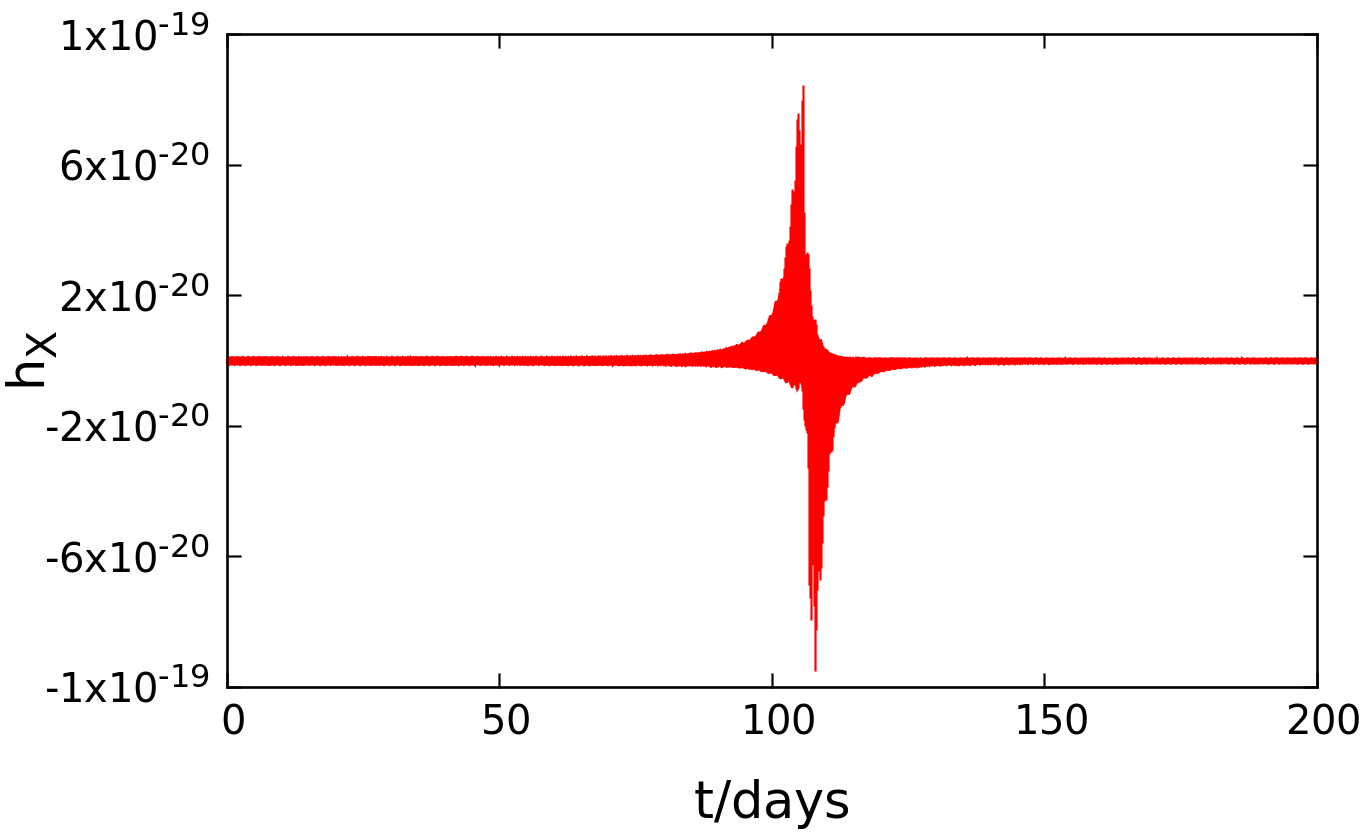}
\caption{\small The top and bottom figures correspond to + and $\times$ polarization, respectively, of the waveform of GWs 
from the hierarchical triplet model IA1 with KL oscillation. The observer distance is 10 {\rm kpc} at inclination $\iota=45^\circ $ measured from the initial outer orbital plane.}
\label{fig:kozaiwaveform}
\end{figure}
\begin{figure}
  \centering
\end{figure}
In Fig. \ref{fig:kozaiwaveform}, assuming the observer distance is 10 kpc, 
we present the waveform of two 
polarization modes for one KL oscillation cycle.

For the small eccentricity until day 98, 
the amplitudes are low as $h\sim 10^{-21}$, but 
when the inner binary is in high eccentricity regime 
around day 100, we find a significant rise in amplitude
as $h\sim 10^{-19}$.

\begin{figure}[h]
 \includegraphics[width=0.70\linewidth]{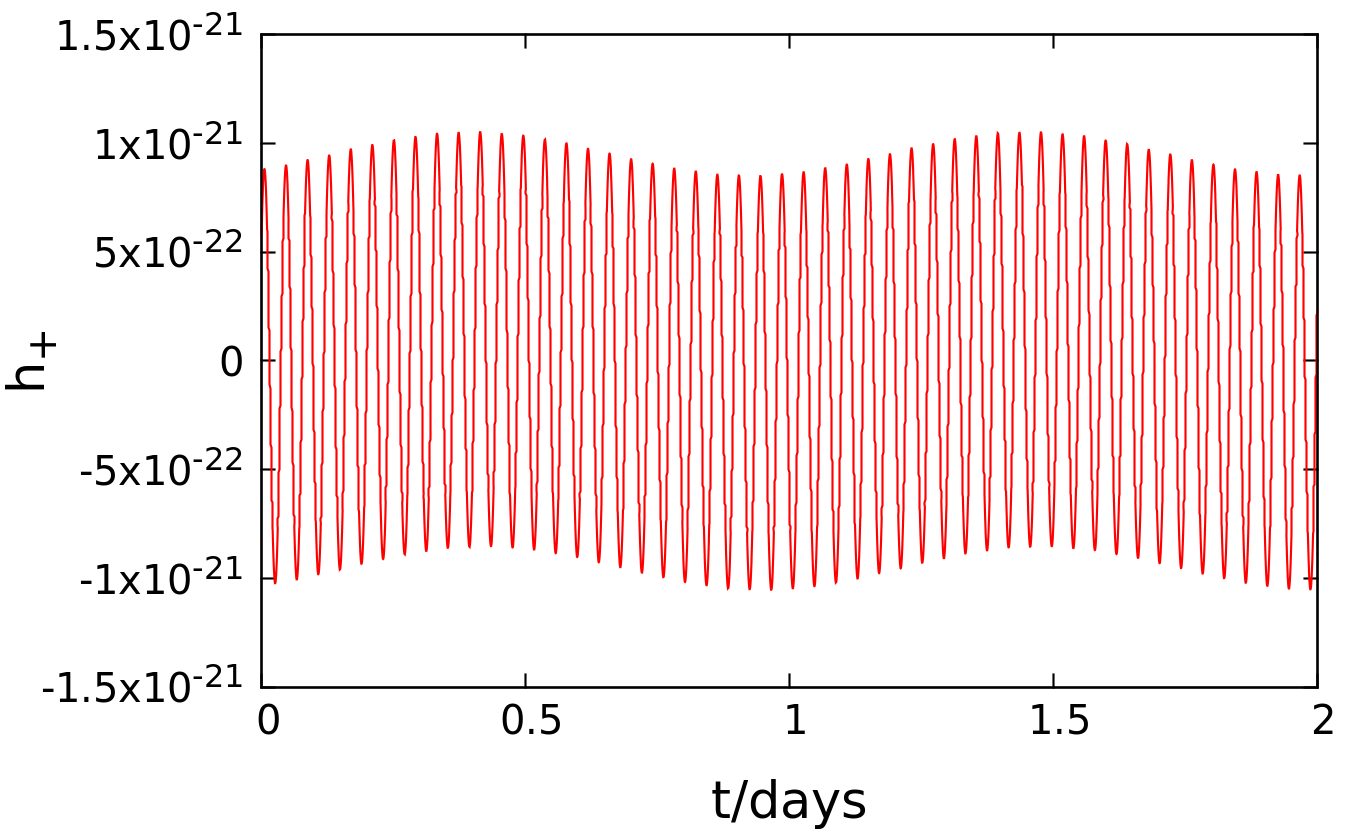}
 \\
\includegraphics[width=0.70\linewidth]{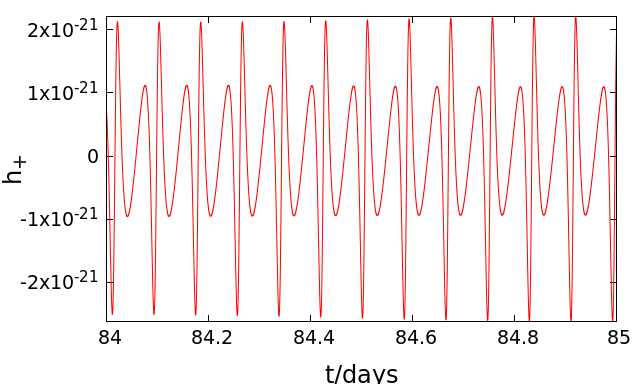}
 \\
 \includegraphics[width=0.70\linewidth]{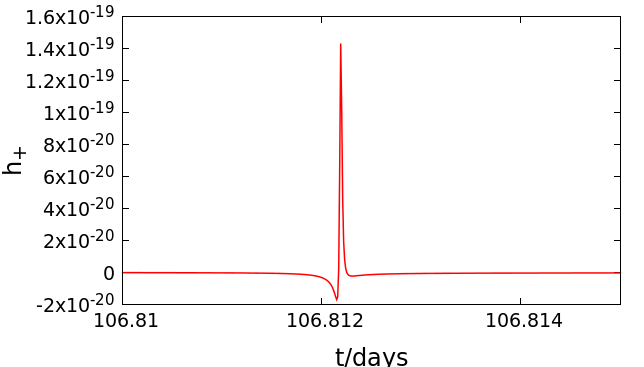}
\caption{\small The zoomed in parts of waveform of $+$ polarization in low and high eccentricity regime. The top figure shows 
the waveform in low eccentricity period (the first two days), 
 while the middle and bottom ones depict the waveforms in the period of
moderate eccentricity ($e_{\rm in}\approx 0.36$) and during periastron passage of high eccentricity ($e_{\rm in}\approx 0.98$),
respectively.}
  \label{fig:kozaiwaveform_2days_+}
\end{figure}

\begin{figure}[h]
  \centering
  \includegraphics[width=0.70\linewidth]{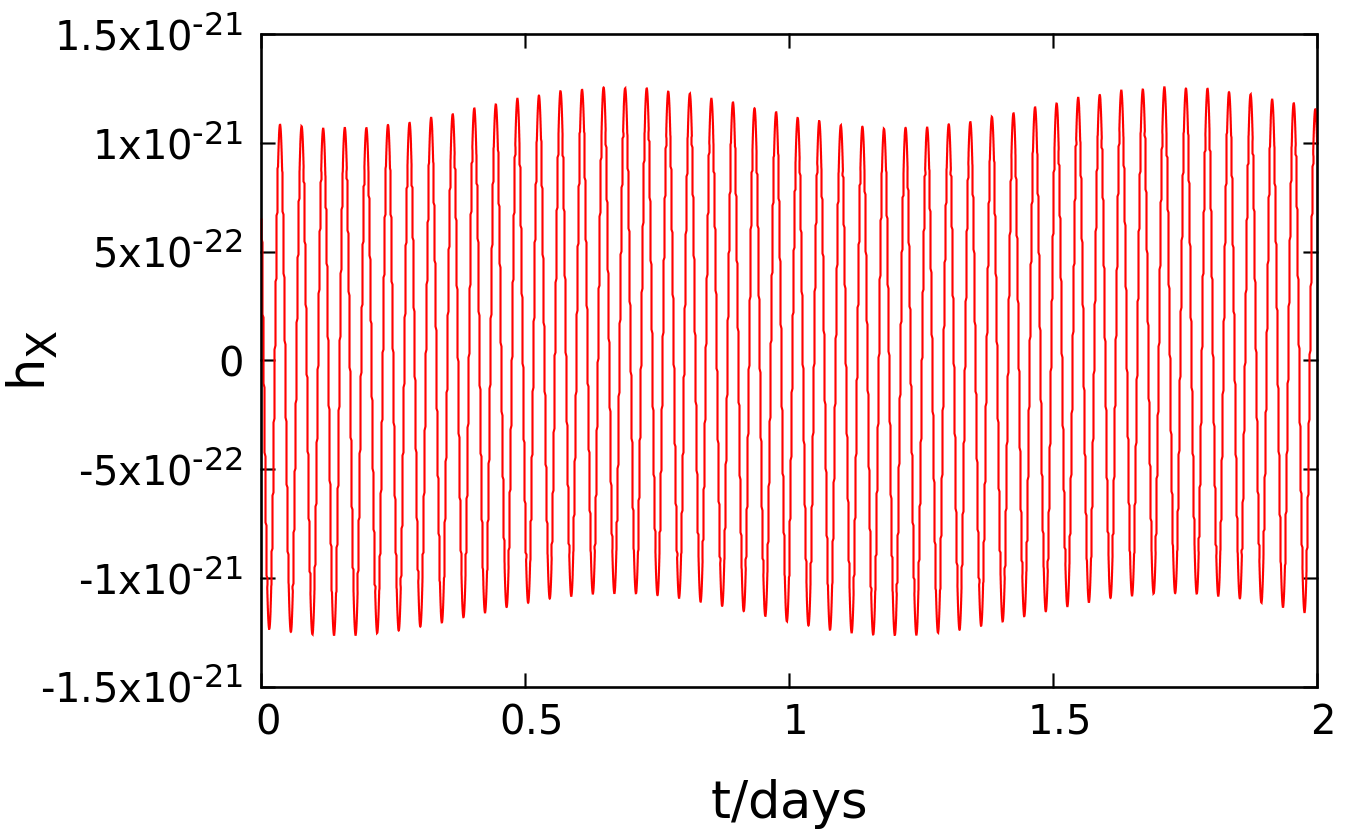}
  \\
 \includegraphics[width=0.70\linewidth]{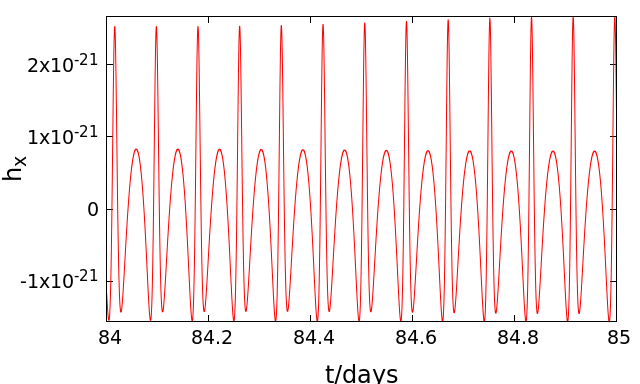}
  \\
\includegraphics[width=0.70\linewidth]{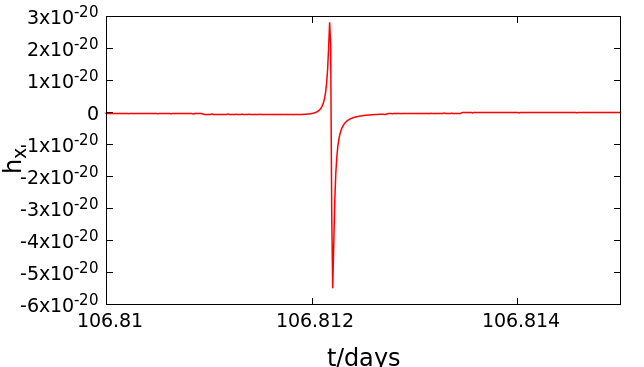}
\caption{\small The same zoomed figure as Fig. \ref{fig:kozaiwaveform_2days_+} for $\times$ polarization.}
  \label{fig:kozaiwaveform_2days_x}
\end{figure}

To see more detail, we enlarge the part of waveforms.
In Fig. \ref{fig:kozaiwaveform_2days_+},
we show the waveform of $+$ polarization mode.
The top figure depicts the data for the first two days, when 
the eccentricity is small, 
while the middle and bottom ones show the data from day 84 to day 85
and one  from day 106.80 to day 106.85, 
respectively, when the eccentricity becomes large.
We also show the similar enlarged waveforms for $\times$ mode in Fig. \ref{fig:kozaiwaveform_2days_x}.

These features can be easily understood when we analyze more simple systems such as an isolated binary or a coplanar motion of a hierarchical triple system, for which the results are summarized in Appendix \ref{waveform_IB_CTS} 
for Models 1A1$_{\rm CC}$, 1A1$_{\rm CE}$, I$_{\rm C}$ and 
I$_{\rm E}$.

For  an isolated circular binary (Model I$_{\rm C}$), the waveform is sinusoidal,
but if the eccentricity is large as $e=0.9$ (Model I$_{\rm E}$), 
the sharp peak in the waveform appears near the periastron point
(see Figs. \ref{fig:isolatedwaveform_c} 
and \ref{fig:isolatedwaveform_e}). The amplitude also 
increases compared with the circular case. 
For a coplanar hierarchical triple system 
(Models 1A1$_{\rm CC}$, 1A1$_{\rm CE}$), 
we find similar features with those in an isolated binary.
When the eccentricity $e_{\rm in}$ is small, we find the sinusoidal waveform, while if $e_{\rm in}$ is large, a sharp peak appears
near the periastron point (see Figs. \ref{fig:coplanarwaveform}
and \ref{fig:ecoplanarwaveform}).
In the case of a triple system, however, when the eccentricity is small,
there appears small modulation in the amplitudes. 
As we explain in Appendix \ref{GW_CTS}, 
we expect two GW sources; inner and outer orbits, and 
the ratio of the amplitudes of GWs from inner and outer orbits
is 
$
{h_{\rm out}/ h_{\rm in}}\sim 0.2
$
for Models IA1$_{\rm CC}$.
The waveform from a hierarchical triple system 
is obtained by superposition of $h_{\rm out}$ and $h_{\rm in}$.
Since the wavelength of GWs by the outer orbit is 
much longer than that by the inner orbit,
the superposition gives
the waveform shown in Fig. \ref{fig:coplanarwaveform}.
The similar modulation is found in Model IA1 when the eccentricity is small
(see top figures in Figs. \ref{fig:kozaiwaveform_2days_+}
and \ref{fig:kozaiwaveform_2days_x}).

When the eccentricity becomes large via KL oscillation, 
the local feature of the waveform is similar to that of 
an eccentric binary or in a coplanar eccentric orbit of a triple system.  
We show the waveforms for two stages in Figs. \ref{fig:kozaiwaveform_2days_+} and \ref{fig:kozaiwaveform_2days_x}: Middle one is in the period of
moderate eccentricity ($e_{\rm in}\approx 0.36$) and the bottom one corresponds to the period during periastron passage of high eccentricity ($e_{\rm in}\approx 0.98$). 
As we will see later, the waveform for the moderate 
eccentricity can be observed when the strain curve just enters
the observable range for LISA, while that for maximum eccentricity 
can be observed when the strain curve reaches maximum.

\begin{figure}[h]
  \centering
   \includegraphics[width=0.7\linewidth]{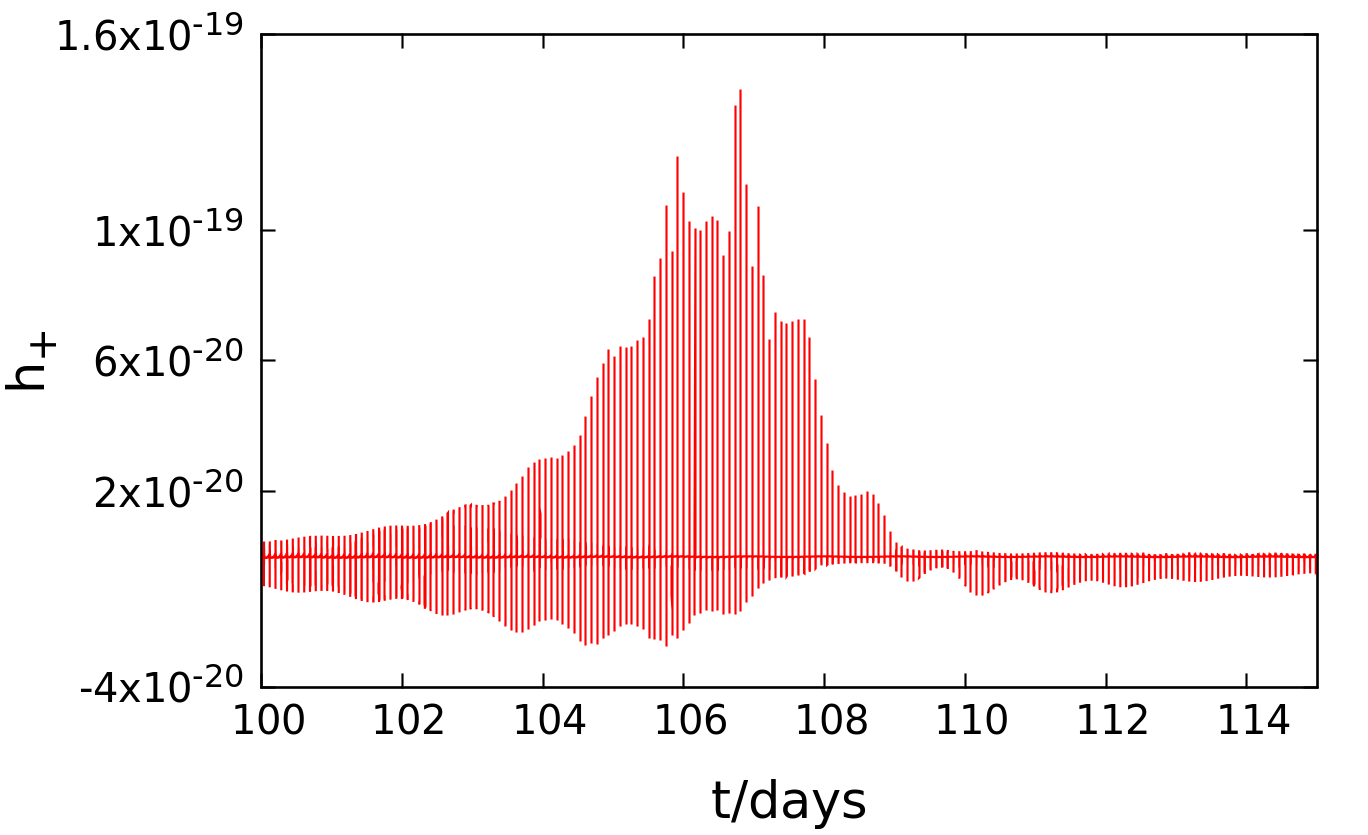}
   \\
   \includegraphics[width=0.7\linewidth]{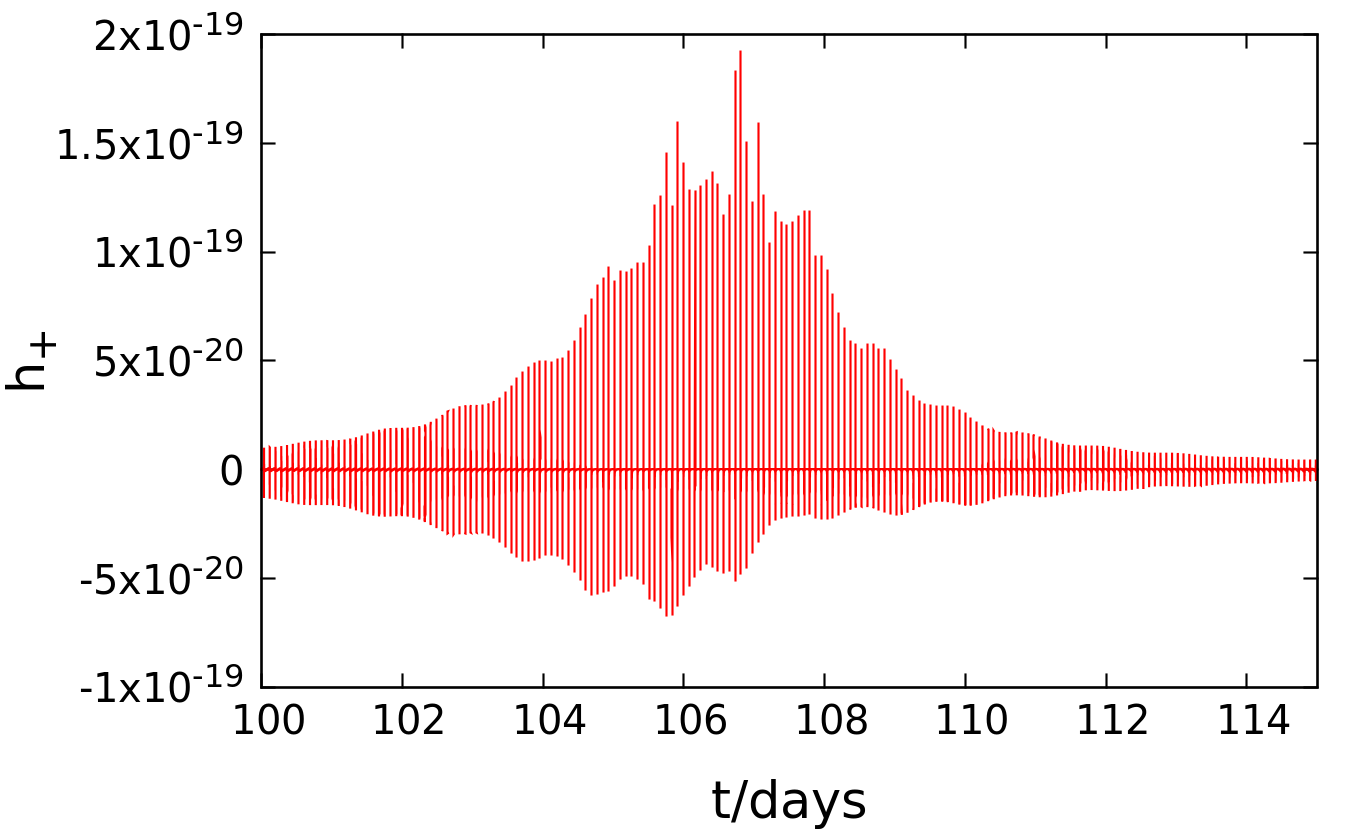}
\caption{\small The global shape of the waveform with $+$ polarization
for 15 days (from day 100 to day 115). It depends on the observer position. The top figure shows
the observer at inclination $45^\circ$, while 
the bottom one is for the observer at inclination $90^\circ$. }
\label{fig:inc_observer_hp}
\end{figure}

\begin{figure}[h]
  \centering
  \includegraphics[width=0.7\linewidth]{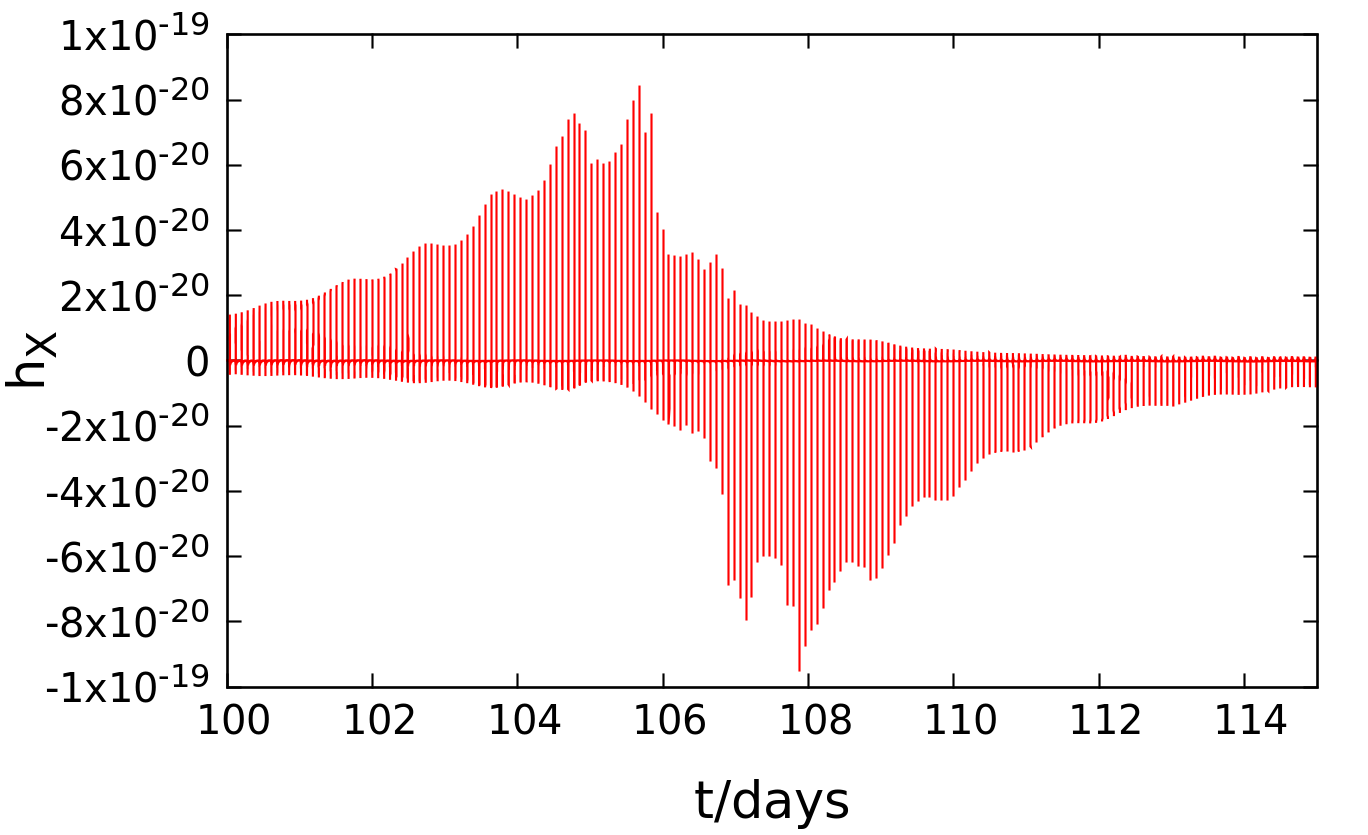}
  \\
    \includegraphics[width=0.7\linewidth]{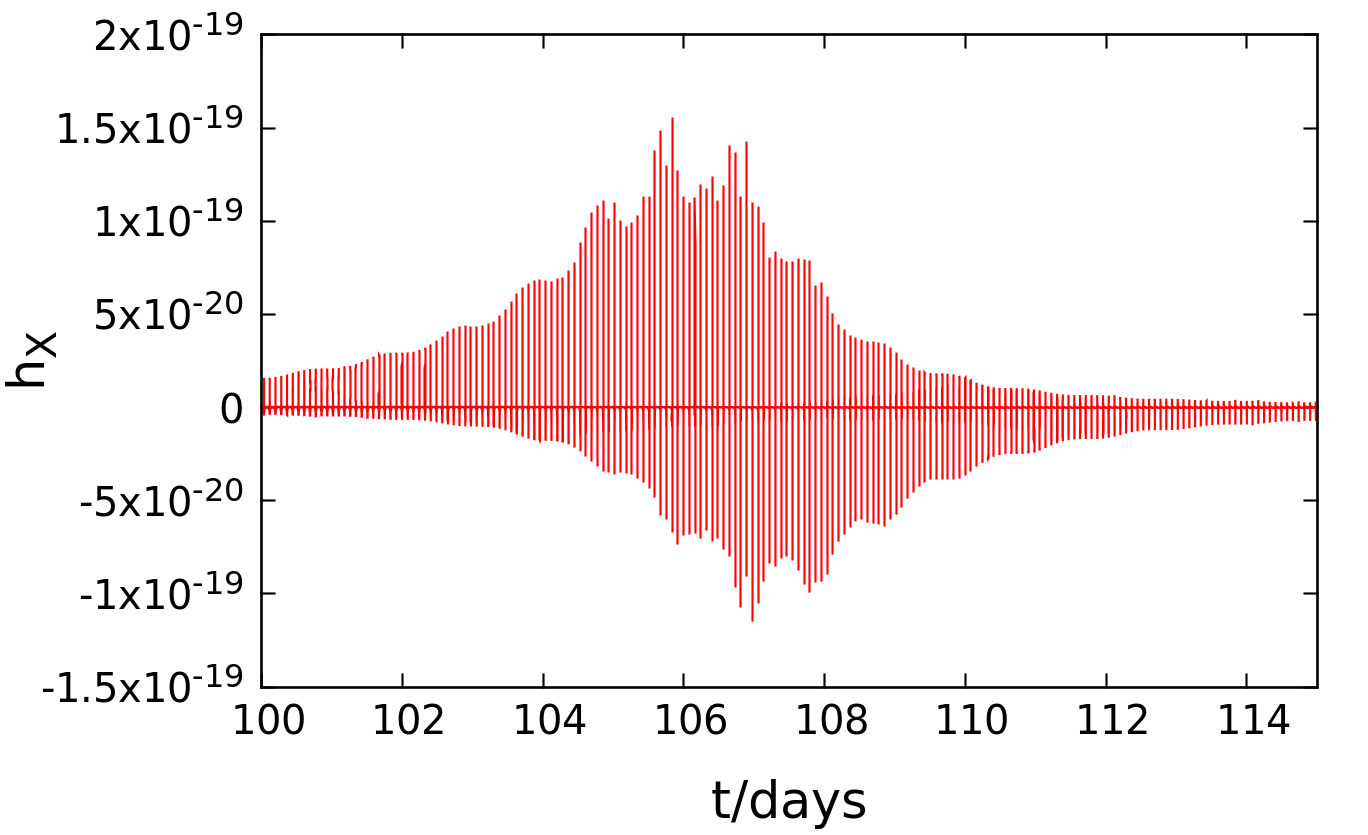}
\caption{\small The same figure as Fig. \ref{fig:inc_observer_hp}
for $\times$ polarization mode.}
\label{fig:inc_observer_hc}
\end{figure}

As shown in Figs. \ref{fig:inc_observer_hp} and \ref{fig:inc_observer_hc}, 
the global feature of the amplitude changes randomly because of the oscillation of the eccentricity, which is different from the 
coplanar eccentric case.

For the observer at a different position, we also find the same 
features. The difference appears in the global shape of the waveform.  We show the examples for the observers at $\iota=45^\circ$
and $\iota=90^\circ$ measured from the initial outer orbital plane
in Figs. \ref{fig:inc_observer_hp} and \ref{fig:inc_observer_hc}.


\subsubsection{\textbf{\underline{GW Spectra}}}
\label{GW Spectra}
Large eccentricities during KL oscillation and variation of eccentricity with time introduce more harmonics in GWs which make the energy spectra rich as compared to the spectra from an isolated binary (shown in Appendix \ref{GW_isolated}). In Fig. ~\ref{fig:kozaispectra}, we show energy spectra for two stages in one KL cycle. During low eccentricity regime, the spectra is similar to that of circular coplanar hierarchical triple (Model IA1$_{\rm CC}$). 

\begin{figure}[h]
  \includegraphics[width=0.70\linewidth]{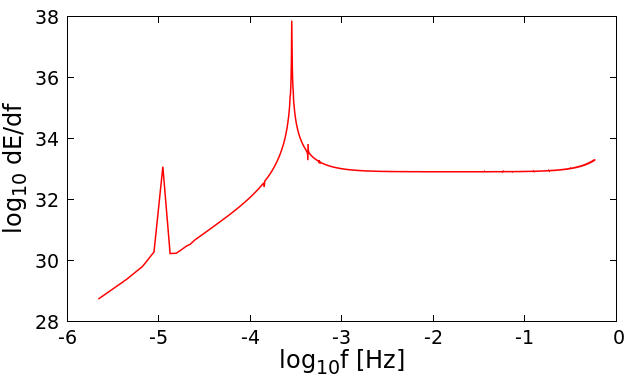}
  \\
  \includegraphics[width=0.70\linewidth]{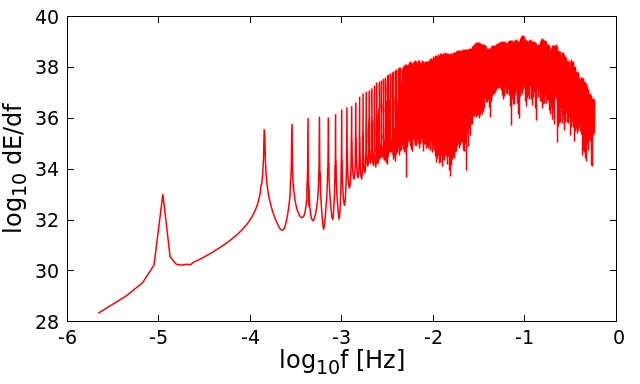}
   \caption{\small The GW energy spectra from the inner binary with KL oscillation for low and high eccentricity regime.
  The top figure shows the spectrum at day 1-5 ($e_{\rm in} \sim 0$), and 
  the bottom one gives that at day 106-108  ($e_{\rm in} = 0.983 \sim 0.987$).
  We find a broad band spectrum for a high eccentric orbit.}
\label{fig:kozaispectra}
\end{figure}

In the top figure in Fig ~\ref{fig:kozaispectra},
which is obtained by the Fourier transformation for the time interval from day 1 to day 5, 
we find two sharp peaks which correspond to 
the two periods of the inner and outer orbits. 
On the other hand, during high eccentricity regime, 
as we see from the bottom one of  Fig ~\ref{fig:kozaispectra},
which is obtained for the time interval 
from day 106 to day 108, 
the spectra is quite similar to the eccentric coplanar 
model IA1$_{CE}$ (see Fig. \ref{fig:spectra_3c}).
Because of high eccentricity, many higher harmonics appear.
As a result, we find a broad band spectrum for a high eccentricity regime.



\subsubsection{\textbf{\underline{Model {\rm IB3}}}}
\label{model BI}
\begin{figure}[h]
  \centering
  \includegraphics[width=0.75\linewidth]{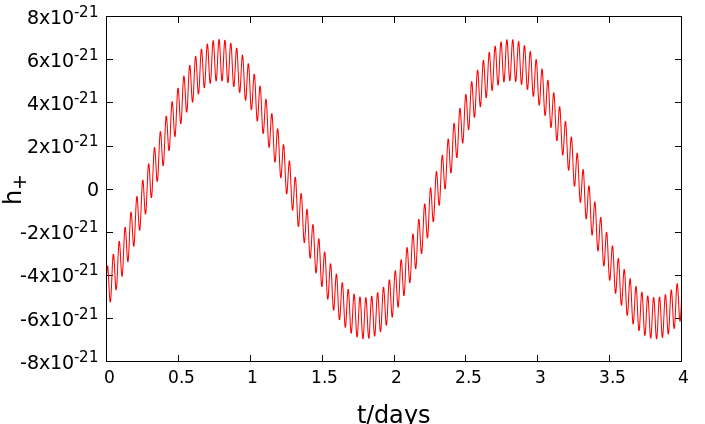}
  \\
\includegraphics[width=0.75\linewidth]{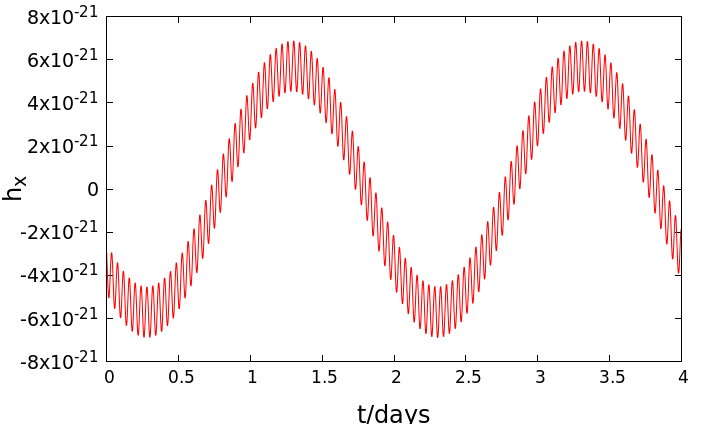}
\caption{\small The same zoomed figures as top ones in Figs. \ref{fig:kozaiwaveform_2days_+} and \ref{fig:kozaiwaveform_2days_x} for $+$ and $\times$ polarizations, respectively.}
  \label{fig:kozaiwaveform_2days_IB3}
\end{figure}

We also perform numerical calculation for Model IB3, i.e.,
$m_1=m_2=10 M_\odot$, $m_3=10^3 M_\odot$, $a_{\rm in}= 0.01 {\rm AU}$, and $a_{\rm out}= 0.5 {\rm AU}$.
We find that the properties of GWs are similar to 
Model IA1.
There appears KL oscillation, in which the maximum eccentricity 
becomes 0.99. 
As for the waveform, when the eccentricity is small,
we also find a sinusoidal shape with modulation.
However, in this case, the amplitude of GWs from the outer orbit is larger than 
that from the inner orbit. 
In fact, the ratio of two amplitudes becomes 
$
{h_{\rm out}/h_{\rm in}}\sim  4,
$
assuming those orbits are almost circular.
Hence the modulation is much larger than the case of Model IA1
as shown in Fig. \ref{fig:kozaiwaveform_2days_IB3}.

While, during the highly eccentric stage,
it shows a sharp peak by passing near 
a periastron point just as that in Model IA1.
The GW spectra also change in one KL cycle 
from ones with two sharp frequencies, corresponding to two orbital frequencies, 
to a broad band spectra.
The difference appears only in the time scales given in Table  ~\ref{tab:ini}.
\subsection{Observability: Frequency, Strain,  and SNR} 

To examine the possibility of observation of the present
triple models by future space-based detectors, we have to check whether 
the frequency and strain are in the observable range 
and evaluate the signal to noise ratio (SNR).

\begin{figure}[h]
    \centering
    \includegraphics[width = 7.5cm]{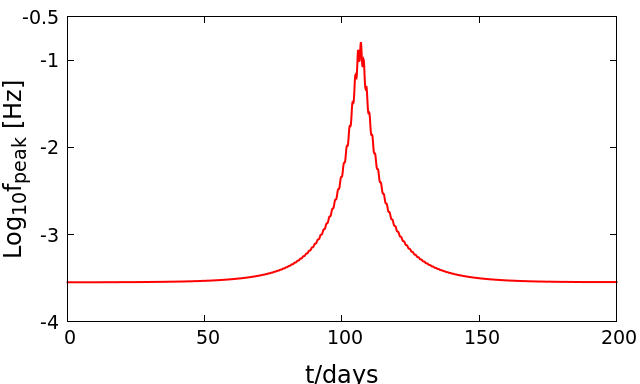}
    \caption{\small Time evolution of the peak frequency for 
    one KL oscillation cycle for Model IA1
    ($m_1=m_2=m_3=10M_\odot$ and $a_{\rm in}=0.01 {\rm AU}, a_{\rm out}=0.1 {\rm AU}$).}
    \label{fig:fpeak}
\end{figure}

We first plot $f_{\rm peak}$ for Model IA1. We show in Fig. ~\ref{fig:fpeak} that due to KL oscillations the frequency sweeps up to the detectable range of proposed future space-based interferometers. 

\begin{figure}[h]
  \centering
   \includegraphics[width=8.5cm]{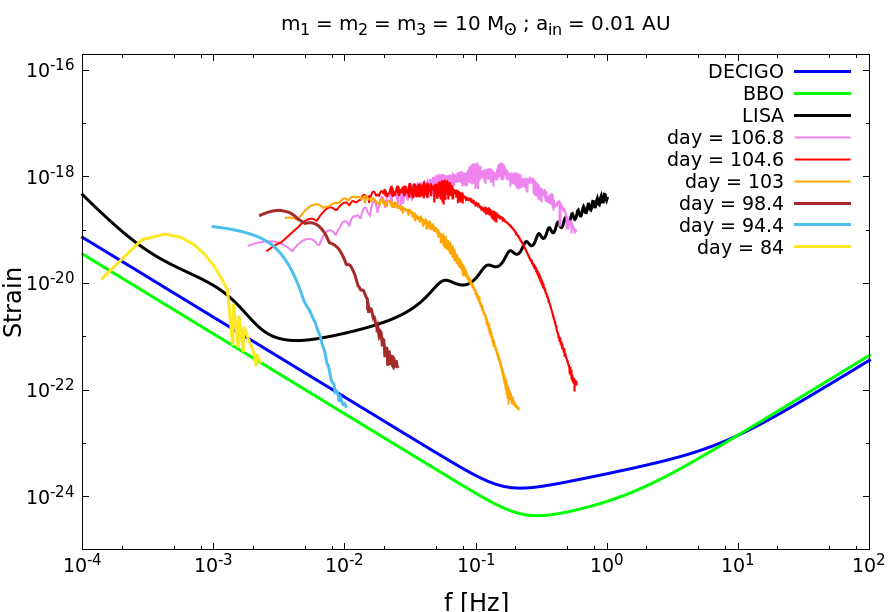}
\caption{\small The strain versus frequency plot for observer at inclination 90$^\circ$. The black, blue,  and green curves show LISA, DECIGO, and BBO design sensitivity. The yellow curve shows the strain curve entering the observable range for LISA  with $e_{\rm in}$ = 0.36 at day 84. The blue, brown, orange, red, and violet curves show the strain evolution with $e_{\rm in}$ = 0.361 at day 84, 0.63 at day 94.4, 0.824 at day 98.4, 0.944 at day 103, 0.972 at day 104.6, and  0.987
(=$e_{\rm in, max}$) 
at day 106.8, respectively.}
\label{fig:hfplot_A}
\end{figure}

Although the GWs from an inspiral phase is expected to be small 
when the eccentricity is small, 
we may have a chance to find a large amount  of GW emission
when the eccentricity becomes large enough via KL oscillations.
 We then plot the strain versus frequency curve for Model IA1 in 
Fig. \ref{fig:hfplot_A}.

The black, blue, green curves correspond to strain sensitivity curves of LISA, DECIGO and BBO respectively \citep{Cornish2018,Seto2017}. We plot GW 
characteristic strain curve for Model IA1 at different days. 
The yellow curve shows 
the strain just entering the LISA sensitivity curve.
It happens at day 84-85 when the binary is moderately eccentric with $e_\mathrm{in} \sim 0.36$.
The light blue, brown, orange, red and  violet curves are 
plotted for 
day 94-95 with $e_\mathrm{in} \sim 0.63$, 
day 98-99 with $e_\mathrm{in} \sim 0.824$, 
day 102-103 with $e_\mathrm{in} \sim 0.944$, 
day 103-105 with $e_\mathrm{in} \sim 0.972$, 
day 106-108 with $e_\mathrm{in} \sim 0.987$, respectively.
The violet curve gives the maximum strain for this model.
After day 108, the eccentricity decreases and then 
the strain curves decrease to 
the yellow one through the red, orange, brown, and light blue ones.
These curves show the evolution of the strain in one KL cycle.
At high eccentricity regime ($e_{\rm in}\gsim 0.36$) of KL cycle, 
the GWs may be observable for about 48 days
every KL cycle ($\sim $ 176 days ) by LISA, DECIGO or BBO.
Since the KL oscillation is repeated, once we find this object,
we have enough time for next observation.

\begin{figure}[h]
  \centering
  \includegraphics[width=8.5cm]{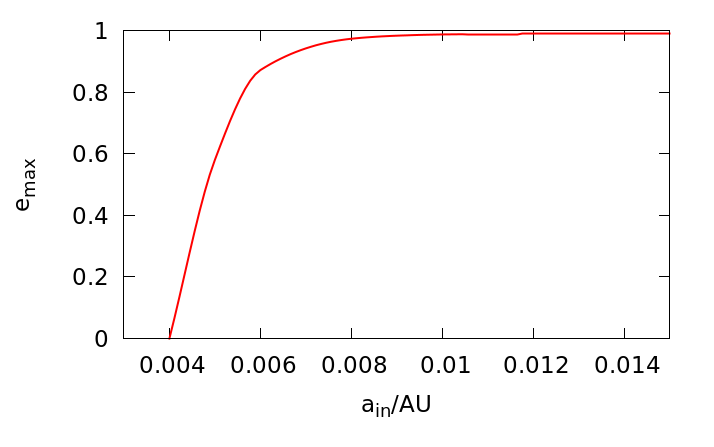}
\caption{\small The maximum eccentricity in terms of $a_{\rm in}$ for model IA1.
We find $e_{\rm in, max}\approx 1$ for $a_{\rm in} \gsim 0.01 {\rm AU}$, while $e_{\rm in, max}$ 
decreases rapidly beyond $a_{\rm in}\sim 0.01 {\rm AU}$.}
\label{fig:ecc_ain}
\end{figure}

One may wonder that the GW signal may be louder when $a_{\rm in}$
becomes smaller via an evolution of the inner binary by 
the emission of GWs.
For example, one may choose $a_{\rm in}=0.006 {\rm AU}$,
which is still in the stable range
for KL oscillations (see Fig. \ref{fig:parameter_ain_case_a}).
However it does not give a louder signal.
It is because 
the maximum eccentricity $e_{\rm in, max}= 0.87$ 
for $a_{\rm in}=0.006 {\rm AU}$  is not 
high enough to give an observable strain 
compared with  Model IA1. 
We show the change of the maximum eccentricity $e_{\rm in, max}$
in terms of $a_{\rm in}$ in Fig. \ref{fig:ecc_ain}.

We find that the maximum eccentricity is very close to unity for
$a_{\rm in}\gsim 0.01 {\rm AU}$, but 
it decreases rapidly past $a_{\rm in}\sim 0.01 {\rm AU}$.
This is because GR effect suppresses the KL mechanism\citep{tey13,naoz13b,liu15}.
As a result, the case with $a_{\rm in}= 0.01 {\rm AU}$,
which shows the shortest semi-major axis as well as the largest eccentricity,
 gives the loudest characteristic strain.

\begin{figure}[h]
  \centering
  \includegraphics[width=8.5cm]{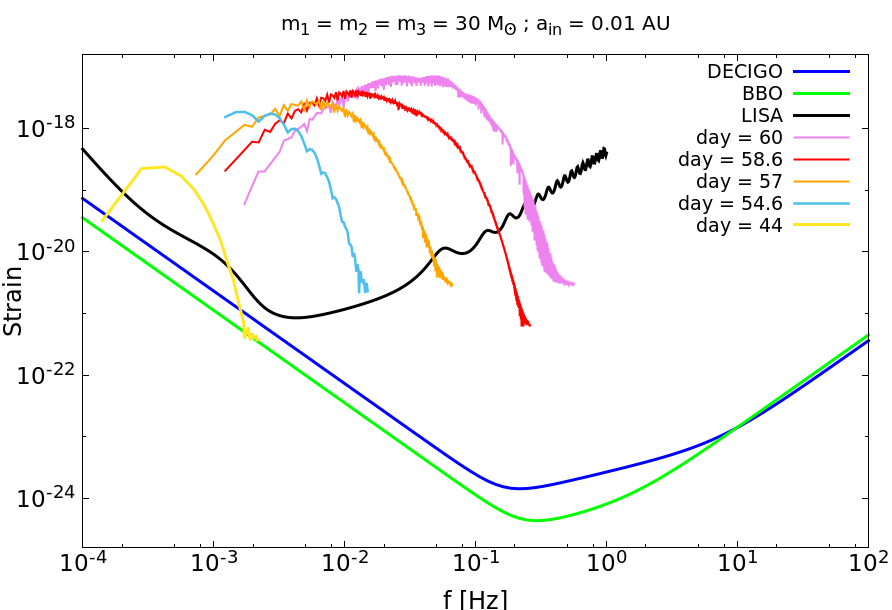}
\caption{\small The strain versus frequency plot for observer at inclination 90$^\circ$. The black, blue, and green  curves show  LISA, DECIGO, and BBO design sensitivity. The yellow curve shows the strain entering LISA sensitivity curve with $e_{\rm in}$ = 0.21 at day 44.
The blue, orange, red and violet curves show the strain evolution with $e_{\rm in}$ = 0.77 at day 54.6, 0.878 at day 57, 0.946 at day 58.6, and  0.964 
(=$e_{\rm in, max}$) at day 60, respectively. The observer distance is 10kpc.}
\label{fig:hfplot_A2}
\end{figure}

In Fig. ~\ref{fig:hfplot_A2}, we also show the similar plot 
for Model IA1$_3$ with $m_1=m_2=m_3 = 30 M_\odot$ 
and $a_\mathrm{in} = 0.01 {\rm AU}$.  When the observer distance is again kept to be at 10kpc,
the signal is louder due to increase in mass.
 We may observe the GWs for about 32 days
every 123 days (one KL cycle) by LISA, DECIGO or BBO.

\begin{figure}[h]
  \centering
  \includegraphics[width=8.5cm]{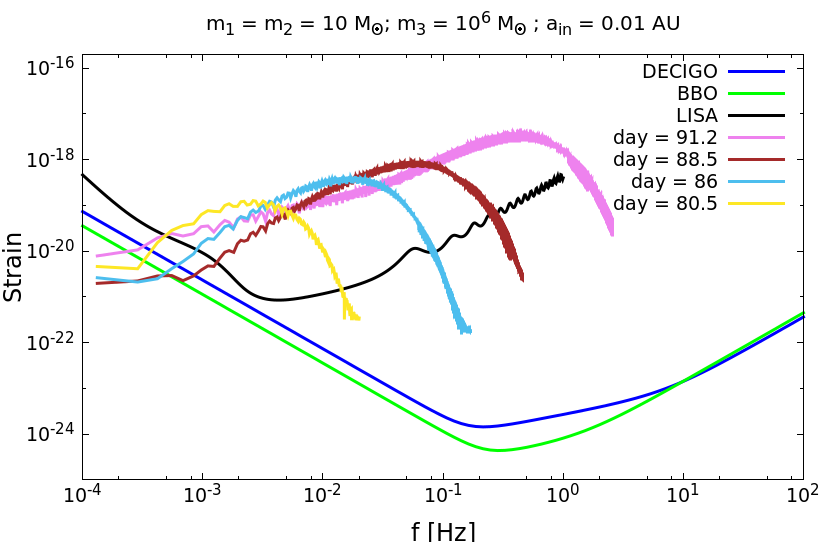}
\caption{\small The same figure  for Model IB6 as Figs. \ref{fig:hfplot_A} and \ref{fig:hfplot_A2}. The black, blue, and  green curves show LISA, DECIGO, and BBO  design sensitivity. The yellow curve shows the strain entering LISA sensitivity curve with $e_{\rm in}$ = 0.80 at day 80.5. The blue, brown and violet curves show the strain evolution with $e_{\rm in}$ = 0.94 at day 86, $e_{\rm in}$ = 0.97 at day 88.5 and 0.99 (=$e_{\rm in, max}$) at day 91.2, respectively. The observer distance is 10 kpc.}
\label{fig:hfplot_B}
\end{figure}

As for Case (B), we give the characteristic strain for Model IB6
in Fig. \ref{fig:hfplot_B}.
The observer distance is kept at 10 kpc.
We also find sufficient large strain for LIGO, DECIGO and BBO.
The observable period is $\sim 32$ days every 96 days (one KL cycle).

A similar model with a SBH binary around SMBH is studied by Hoang et. al with a highly eccentric outer orbit\cite{Hoang2019}.
 They studied the model with $m_1=20 M_\odot, m_2=30 M_\odot$ and $m_3=4\times 10^6 M_\odot$. The semi-major axes are chosen as $a_{\rm in}=0.15{\rm AU}$ and $a_{\rm out}=250{\rm AU}$. 
Since the semi-major axis of the inner binary is larger than our model, the characteristic strain is smaller than our case.
However, because the semi-major axis of the outer orbit is large,
the KL timescale is about $t_{\rm KL}\sim 10$ yrs, and then 
the period of high eccentricity becomes much longer than the orbital period of the inner binary. As a result, we can accumulate the GW data for a longer observational time and then get the high SNR for the observation.

\begin{figure}[h]
  \centering
    \includegraphics[width=8cm]{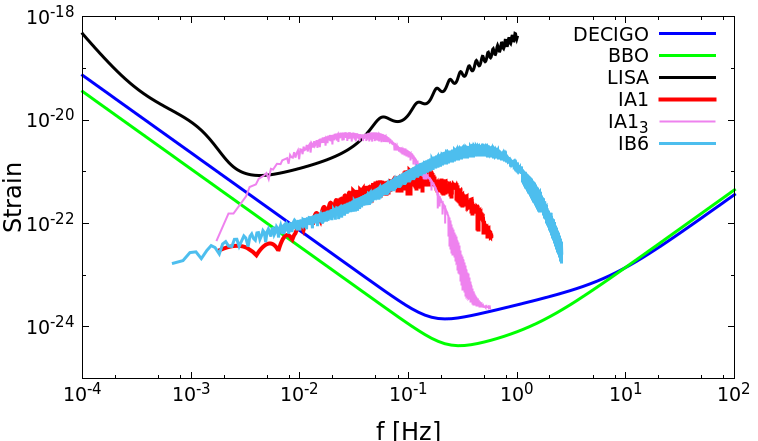}
\caption{\small The strain versus frequency plot for Models IA1, IA1$_3$ and IB6. The black, blue, and green  curves show  LISA, DECIGO and BBO design sensitivities.  The observer distance 
is 16 Mpc.}
\label{fig:hfplot_16Mpc}
\end{figure}

To see the possibility of observation for
the extra-galactic GW source, we also analyze 
the characteristic strain for the observer distance at 16 Mpc,
which is shown in Fig. \ref{fig:hfplot_16Mpc}.

The plot of the frequency vs the characteristic strain gives helpful assessment of detectability provided high signal to noise ratio (SNR) of the waveform. 
Our hierarchical triple system may be one of many source populations captured by future GW space detectors with following merger phase captured by ground detectors.

The area between the source and detector curves is related to SNR by the following relation \citep{Berry2014},

\begin{equation}
({\rm SNR})^2 = \int_{f_{\rm min}}^{f_{\rm max}} d (\log f) \bigg(\frac{h_c(f)}{h_d(f)}\bigg)^2 
\,,
\end{equation} 
where $h_c(f)$ is the strain amplitude  of the source and 
$h_d(f)$ denotes noise amplitude of detector. 
For the models shown in our analysis,
the SNR  for LISA is about $\sim 162$, 210 and 195
for Model IA1, Model IA1$_3$ and Model IB6 at the distance 10kpc, respectively. Since the design sensitivity of BBO and DECIGO is better than LISA, the SNR would be larger for them. 
If the observer is at 16Mpc, the strain amplitude 
for the above models is not loud enough for LISA, but still 
in observable range for DECIGO or BBO.

As for accumulation of data,
we may introduce the following number $N$
\bea
N={\Delta t_{\rm KL}\over P_{\rm in}}\approx \Delta {m_{\rm in}\over m_3}\left({a_{\rm out}\over a_{\rm in}}\right)^3
(1-e_{\rm out}^2)^{3/2}
\label{N_accumulate}
\ena
where $\Delta $ is the fraction of the KL cycle $t_{\rm KL}$ 
when the eccentricity is large enough (e.g. $e_{\rm in}\sim 0.9$).  
We may expect $\Delta \sim 10^{-2}$\cite{Hoang2019}.
If $N$ is large enough, 
we can accumulate the data and then get high SNR
just as the model by  Hoang et. al.\cite{Hoang2019} 
(Note that $N=192$ for their model).
We show the number $N$ for our models in Table \ref{model_and_observability}
in Appendix \ref{Observability of other models}.
Although those values are not so large, 
the accumulation of data may provide larger SNR.
Since the accumulation increases the SNR, it is important to 
study in our models with larger semi-major axes too.


\section{Concluding Remarks}
\label{sec:Conc}
Analysing many models in wide range of mass parameters
with KL oscillations, 
we have shown features of gravitational waves from a hierarchical triple system.
The waveform changes its shape in time because of oscillation of 
the eccentricity.
When the eccentricity is small, 
the waveform is a sinusoidal shape modulated by a tertiary 
companion,
while it becomes one with sharp peaks 
near periastron point when the eccentricity gets large.

We have also examined the time 
 variation of the characteristic strain curve, which 
 may appear in the observable range of detectors when the eccentricity becomes large
 via the KL oscillations.
Once we find this GWs event, it will repeat every KL oscillation cycle.
It can be the first direct observation of KL oscillation which may further 
 shed light on binary formation channels and surrounding environment of the binary system.

 The interesting remaining questions are 
what happens when the system evolves beyond the KL stable region.
Because of the GW emission, the semi-major axis of the inner binary 
will decrease. Since other parameters such as  $m_1, m_2, m_3$ and $a_{\rm out}$
are conserved, the system will evolve vertically in Figs. \ref{fig:parameter_ain_case_a}
 and \ref{fig:parameter_ain_case_b} or horizontally 
in Figs. \ref{fig:parameter_case_b} and \ref{fig:parameter_case_b_10^9}.
Hence, we reach at the boundary of relativistic periastron shift.
Beyond this boundary, KL oscillation may be suppressed, and then 
the effect of the tertiary component may become weak.
As a result, we expect that GWs may not be observed before
the coalescence of the inner binary.  The question is whether the orbit of the inner binary
is circularized via the GW emission at the coalescence or not.
If the eccentricity remains then, we find the difference from a simple isolated binary system.
We may need numerical relativity or some other approach such as ``effective one body system"
to clarify it.  Note that Huerta et. al. 
recently introduced a method to extract observable signatures 
from moderately eccentric binaries by using Numerical Relativity templates \citep{huerta19}.

Another interesting issue is what happens when the LT effect becomes important.
It can be the case if we consider SMBH.
In Fig.  \ref{fig:parameter_case_b_10^9}, if the total mass of the inner binary 
is smaller than $10^3 M_\odot$, when the semi-major axis decreases,
the system will evolve into the region where the  LT precession becomes important. 
In this region, we may find a chaotic KL oscillation\citep{Wang19}, which may 
provide us strong emission of GWs. In order to study such a process, we have to include 
higher PN effects, i.e., 
at least 1.5 PN terms. 

The work on these two issues is in progress.
As for a triple system, we are also interested in more 
general situations, in which we expect chaotic behavior of 
a system. The GWs from such a chaotic system should also be 
studied. Since it may be difficult to make templates, 
we should find some other ways to clarify typical features
of such a system
in GW observations \citep{Suzuki:1999si,Kiuchi:2004bv,Kiuchi:2007ue}.

 One of the most important issues about the GWs from a hierarchical triple system is the event rate.
 As for a formation of three body system, there are many works including numerical 
 $N$-body simulation \citep{Glutekin2006,Samsing2014,Fabio2016}.
 The event rate depends on many uncertain factors like star formation rate, 
 binary fraction and distribution of initial parameters.
Several papers discuss the increase in the merger rate of compact binaries around SMBHs due to KL oscillations \citep{Antonini2012,hoang18,trani19,Hoang2019}. The binaries in these triple systems undergo several KL cycles finally leading to the merger. 
In Ref. \citep{Antonini2012}, the authors estimate the rate to be 
0.56 ${\rm Gpc}^{-3} {\rm yr}^{-1}$. Similar analysis is done by Refs. \citep{hoang18,trani19}, estimating the rate ~ 1-3 ${\rm Gpc}^{-3}{\rm yr}^{-1}$. 
The difference is caused by different distance regimes from the SMBH and 
different initial conditions. For isolated triples, 
the merger rate to be 6 ${\rm Gpc}^{-3} {\rm yr}^{-1}$ in the absence of natal kicks\citep{Tremaine17}. Globular clusters may harbor IMBHs at their centers. 
Ref. \citep{Omer19} studies the triple systems composed of the Stellar-mass binary with IMBH as a third body. The merger event rate induced by KL oscillations is estimated to be  
0.06-0.46 ${\rm Gpc}^{-3} {\rm yr}^{-1}$.

We are interested in the models where merger timescale is much longer than Kozai timescale. Hence, if there are many KL oscillations before the merger, the signal about KL oscillations may lie in the observable range.  In Ref. \citep{Lisa2019}, using double-averaged equations, they analyze the fraction of KL binaries in two channels (isolated triples and galactic center) lying in the LISA frequency range. They show that of all merging binaries considered in their simulation, $\sim 39\%$ of the isolated-triple channel, and $\sim 22\%$ of the galactic-center channel, display significant KL oscillations in LISA frequency band.

However, the present models of a hierarchical triple 
system with KL oscillations, 
which gives observable GWs in an inspiral phase, may be
difficult to be formed by evolution of a triple star system 
 because both semi-major axes 
are very short. 
One possible way to form such a system would be a capture process. 
 The third body is bounded due to a close encounter with a BH binary in globular clusters or near galactic centers.
 The question is how likely such an interaction is. 
 In \citep{Samsing2014,zevin19}, binary-single and binary-binary BH interactions are numerically simulated and several possible end states are discussed. 
In Ref. \citep{zevin19}, they have shown that stable triple formation end state has higher cross section compared to some other possible end states.

 We are planning to study in detail the formation process and probability to evaluate the event rate of compact hierarchical triples such that KL oscillations are not quenched by relativistic effects.\\

\section*{Acknowledgements}
We would like to acknowledge Gungwon Kang, Seiji Kawamura, Kaye Li, Naoki Seto, Takahiro Tanaka, and Alessandro A. Trani for useful discussions.
P.G. is supported by Japanese Government (MEXT)
Scholarship.
This work was supported in part by JSPS KAKENHI Grant Numbers   
JP17H06359 and JP19K03857, and by Waseda University Grant for Special
Research Projects (Project number: 2019C-254 and 2019C-640). 

\bibliography{GWHTSKL}

~
\newpage
\begin{widetext}

\appendix
\section{Waveform and Spectra for an Isolated Binary and 
a Coplanar Triple System}
\label{waveform_IB_CTS}
\subsection{GWs from an isolated binary}
\label{GW_isolated}
First, we consider an isolated binary system for the illustration of waveform from a two-body configuration. 
In  Fig.~\ref{fig:isolatedwaveform_c}, we show gravitational waveform from circular binary with frequency twice the orbital frequency, $2 f_{\rm orb}$. 
\begin{figure}[h]
  \centering
  \includegraphics[width=0.35\linewidth]{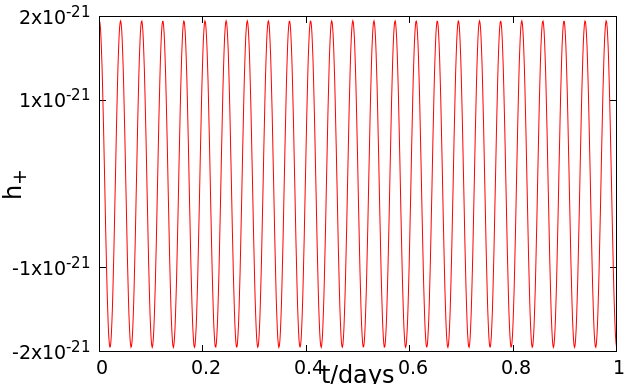}
    \hskip 1cm
\includegraphics[width=0.35\linewidth]{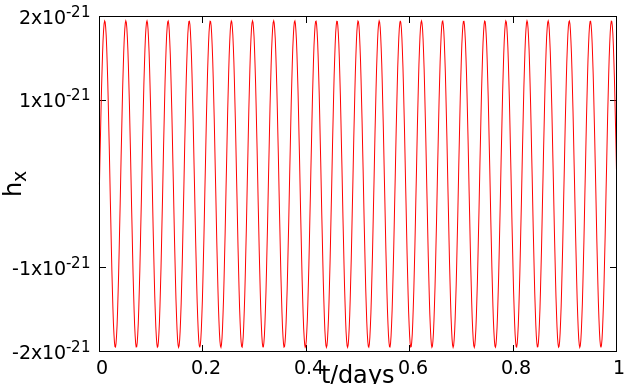}
 \caption{\small The waveform from an isolated circular binary for Model IC ($m_1=m_2=10 M_\odot$ and $a_{\rm in}=0.01 {\rm AU}$). The observer distance is 10{\rm kpc} (Face on). The left and right figures correspond to + and $\times$ polarization, respectively.}
\label{fig:isolatedwaveform_c}
\end{figure}

When a binary has an elliptical orbit, 
GW signal is modulated due to different speeds in a single orbit. It results in 'spikes' due to higher emission near the periastron and hence higher amplitude as shown in Fig.~\ref{fig:isolatedwaveform_e}.
\begin{figure}[h]
  \includegraphics[width=0.35\linewidth]{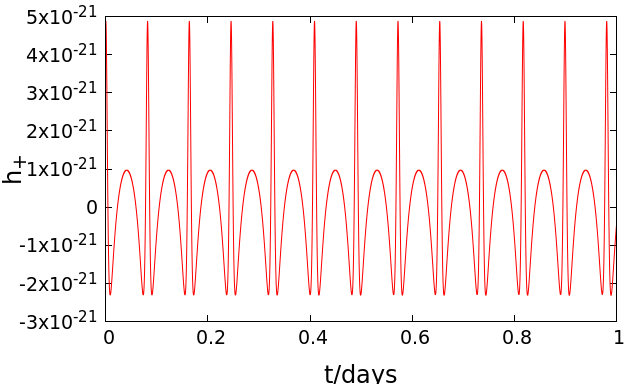}
   \hskip 1cm
 \includegraphics[width=0.35\linewidth]{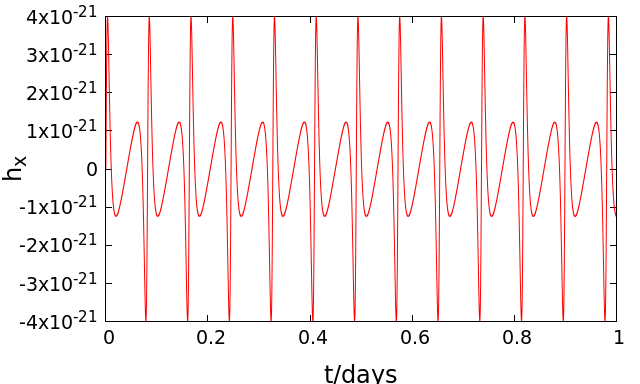}
\caption{\small The same figure as Fig. ~\ref{fig:isolatedwaveform_c}
for an isolated eccentric binary. }
\label{fig:isolatedwaveform_e}
\end{figure}

The energy spectra for Models I$_{\rm C}$ and I$_{\rm E}$ are shown in Fig. ~\ref{fig:spectra_binary}.
As expected, the spectra from an isolated circular binary shows one peak corresponding to twice of the orbital frequency. For the eccentric orbit, we find that 
many higher harmonics are produced because of high eccentricity.
\begin{figure}[h]
 \includegraphics[width=0.35\linewidth]{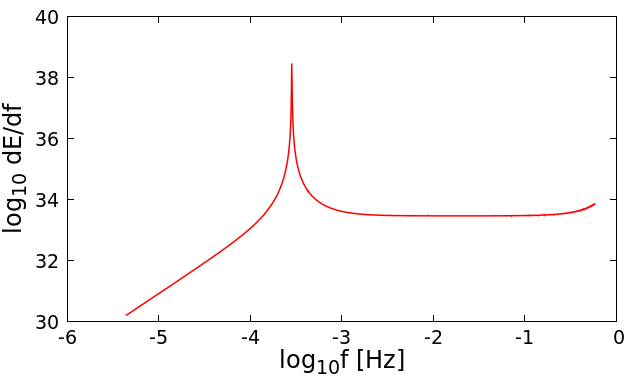}
   \hskip 1cm
\includegraphics[width=0.35\linewidth]{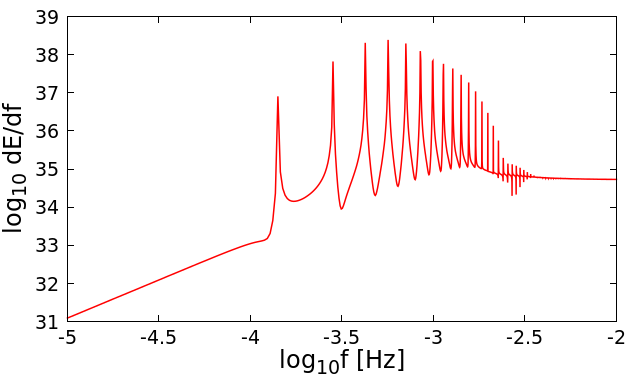}
 \caption{\small The energy spectra for Model I$_{\rm C}$ (a circular binary) and I$_{\rm E}$ (an eccentric binary) with $m_1=m_2=10 M_\odot$ and $a_{\rm in}=0.01 {\rm AU}$. Many higher harmonics are produced because of high eccentricity.}
 \label{fig:spectra_binary}
\end{figure}

\subsection{GWs from a coplanar triple system}
\label{GW_CTS}
Here, we present the results from coplanar systems with $m_1=m_2=m_3=10 M_\odot$ and $a_{\rm in}=0.01 {\rm AU} \,,a_{\rm out}=0.1 {\rm AU}$
for both circular and eccentric inner binary, i.e., Models IA1$_{\rm CC}$ and IA1$_{\rm CE}$.

The waveform  from a circular binary in a coplanar hierarchical triple shows two harmonics corresponding to twice of inner and outer orbital frequencies.
As the case of a circular isolated binary, 
we also find a sinusoidal oscillation (see Fig.~\ref{fig:coplanarwaveform}).
However, there are two sources of GWs; inner and outer orbits. 
Evaluating the amplitudes of GWs from inner and outer orbits by Eq. (\ref{amplitude_binary}), 
the ratio of those wave amplitudes is given by
\bea
{h_{\rm out}\over h_{\rm in}}\sim {(m_1+m_2)m_3\over m_1 m_2}{a_{\rm in}\over a_{\rm out}}\,,
\label{hout_hin}
\ena
which value is $\sim 0.2$ for Models IA1$_{\rm CC}$.
It means that there exists 20\% contribution in the amplitude 
from the outer orbit. 
The waveform from a hierarchical triple system 
is obtained by superposition of those GWs with 
twice of the inner orbit frequency $2f_{\rm in}$
and the outer one $2f_{\rm out}$.
Since the wave length of GWs by the outer orbit is 
much longer than that by the inner orbit,
the superposition gives
modulation shown in Fig. \ref{fig:coplanarwaveform}.

Fig. ~\ref{fig:ecoplanarwaveform} shows waveform 
from an eccentric inner orbit with $e_{\rm in}=0.9$
in a hierarchical coplanar system.
In this case, the amplitude becomes large near the periastron point 
just as the case of an eccentric isolated binary.

\begin{figure}[h]
  \includegraphics[width=0.35\linewidth]{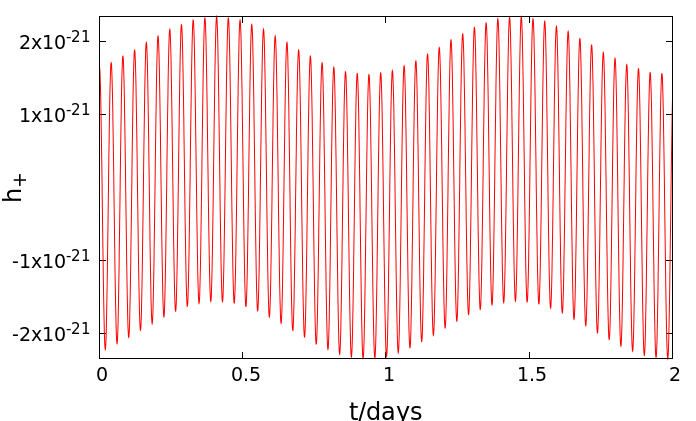}
  \hskip 1cm
  \includegraphics[width=0.35\linewidth]{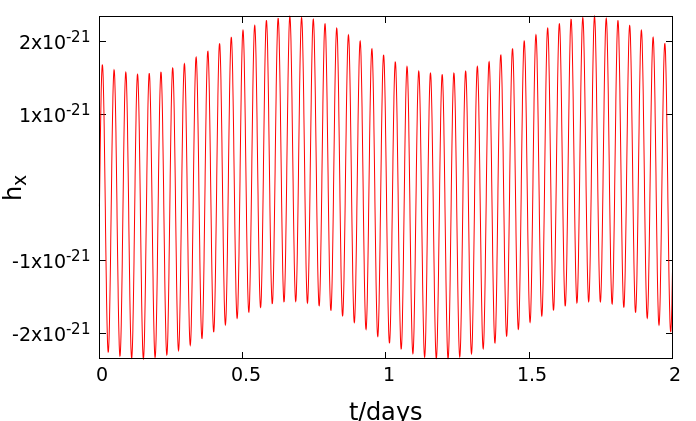}
\caption{\small The left and right figures correspond to + and $\times$ polarization of GWs for Model IA1$_{\rm CC}$
(a coplanar circular binary  in a hierarchical triple system
with $m_1=m_2=m_3=10 M_\odot$ and $a_{\rm in}=0.01 {\rm AU} \,,a_{\rm out}=0.1 {\rm AU}$), respectively.
The observer distance is 10{\rm kpc} (Face on).}
\label{fig:coplanarwaveform}
\end{figure}

\begin{figure}[h]
   \includegraphics[width=0.35\linewidth]{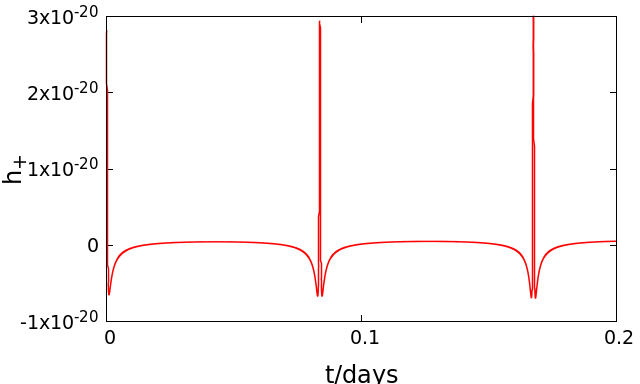}
    \hskip 1cm
 \includegraphics[width=0.35\linewidth]{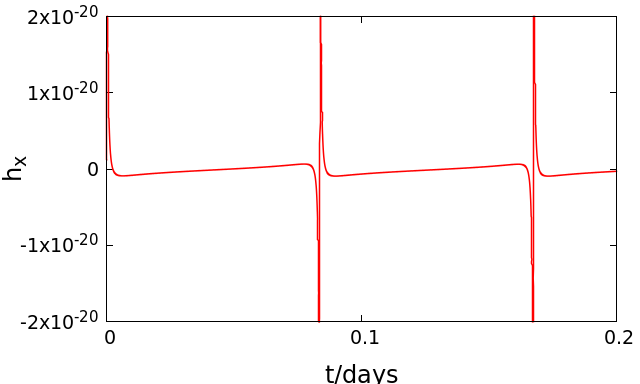}
\caption{\small The same figure are Fig. \ref{fig:coplanarwaveform} 
for Model IA1$_{\rm CE}$ (a coplanar eccentric 
binary in a triple system).
}
\label{fig:ecoplanarwaveform}
\end{figure}
\begin{figure}
  \includegraphics[width=0.34\linewidth]{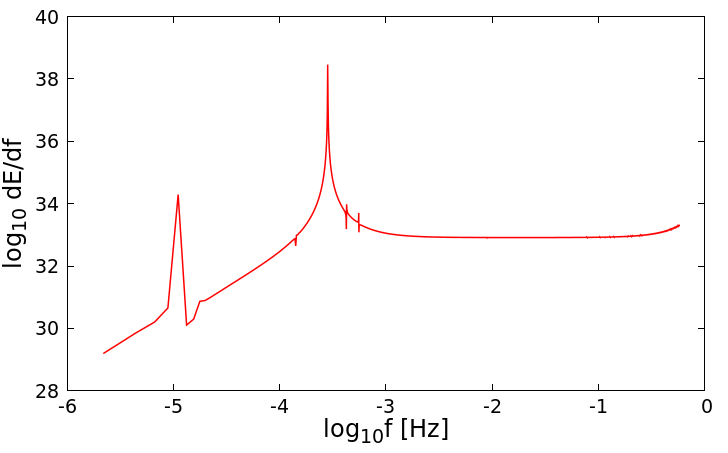}
   \hskip 1cm
 \includegraphics[width=0.34\linewidth]{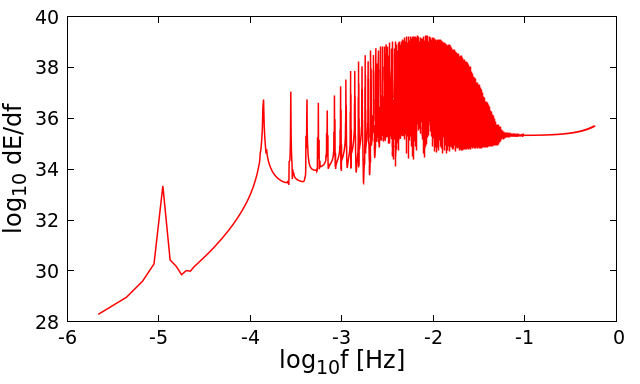}
\caption{\small The energy spectra of a circular coplanar (left [Model IA1$_{\rm CC}$]) and an eccentric coplanar binary (right [Model IA1$_{\rm CE}$]) in a hierarchical triple system, respectively.} 
\label{fig:spectra_3c}
\end{figure}

The energy spectra for Models IA1$_{\rm CC}$ and IA1$_{\rm CE}$ are shown in Fig. \ref{fig:spectra_3c}. 
Just as the case of an isolated circular binary,
the spectrum from a circular inner binary in a coplanar hierarchical system shows two sharp
peaks corresponding to twice  the orbital frequencies of outer and inner orbits. 
This confirms that the GWs from a coplanar hierarchical system is given by
the superposition of two waves from inner and outer orbits.

Unlike circular orbits,
many higher harmonics appear 
in the spectra from an eccentric inner binary in a coplanar hierarchical system.
 As a result, we find a broad band spectrum 
of GWs, which is a little different from the spectrum 
of the eccentric isolated binary. 
This is because the eccentricity 
in a hierarchical triple system is oscillating, 
which gives a rich structure in a spectrum.

\section{Model Summary and Observability of Other Models}
\label{Observability of other models}
In Table \ref{model_and_observability}, we summarize the parameters of our models.
As for the semi-major axes, $a_{\rm in}$ and $a_{\rm out}$, we have chosen them to find the stable KL oscillations. $P_{\rm in}$ and $P_{\rm out}$ are given by the formulas (\ref{Pin}) and (\ref{Pout}), while 
$t_{\rm KL}$
is evaluated by the time interval between first and second peaks of 
the inner eccentricity in our numerical calculation,
 except for Models IV6, VA6, VIA6, for which we use the formula (\ref{tKL}).

\begin{table}[H]
\begin{center}
  \begin{tabular}{|c||c|c|c||c|c||c|c||c||c||c||}
\hline 
&&&&&&&&&&
\\[-.5em]
Model&$m_1$&$m_2$&$m_3$&$a_{\rm in}$&$a_{\rm out}$&$P_{\rm in}$&$P_{\rm out}$&$t_{\rm KL}$
&$e_{\rm in, max}$&$N$   \\
&$[M_\odot]$&$[M_\odot]$&$[M_\odot]$&$[{\rm AU}]$&
$[{\rm AU}]$&[days]&[days]&[days]& & 
\\[.6em]
\hline 
IA1&$10 $&$10  $&$10  $&0.01&0.1&$0.082$&$ 2.10$&$\sim 176$&$\sim 0.98$&20
\\[.1em]
\hline 
IA1$_3$&$30 $&$30  $&$30  $&0.01&0.1&$0.047$&$ 1.22$&$\sim 123$ &$\sim 0.96$&20
\\[.1em]
\hline
IB3&$10 $&$10  $&$10^3  $&0.01&0.5&
$0.082$&$4.04$&
$\sim 255$
&$\sim 0.99$&25
\\[.1em]
\hline
IB6&$10  $&$10  $&$10^6  $&0.01&5&$0.082$&$ 4.08$&$\sim 96$ &$\sim 0.99 $&25
\\[.1em]
\hline 
\hline 
IIA3&$10  $&$10^3  $&$10^3  $&0.12&1&$0.478$
&$8.15$&$\sim 158$
&$\sim 0.99$&5.8
\\
\hline
IIB6&$10  $&$10^3  $&$10^6$&0.12&10&$0.478$
&$11.5$&$\sim 185$ 
&$\sim 0.99$&5.8
\\[.1em]
\hline
IIIA3&$10^3  $&$10^3  $&$10^3$&0.15&1&$0.474$
&$6.67$&$\sim 226$
&$\sim 0.97$&5.9
\\
\hline
IIIB6&$10^3  $&$10^3  $&$10^6 $&0.15&10&$0.474$
&$11.5$&$\sim 142$ 
&$\sim 0.99$&5.9
\\[.1em]
\hline 
\hline 
IVA6&$10  $&$10^6  $&$10^6  $&15&100&$21.2$
&$258$&$17$ [yrs]
&-&-
\\[.1em]
\hline
VA6&$10^3  $&$10^6  $&$10^6  $&15&100&$21.2$&$ 258$&$17$ [yrs] &-&-
\\[.1em]
\hline
VIA6&$10^6  $&$10^6  $&$10^6  $&$10^2$&$10^3$&$ 258$&$ 6669$  &$ 
1400$ [yrs] &-&-
\\[.1em]
\hline 
\hline 
  \end{tabular}
    \caption{The parameters of our models. We use the formula (\ref{Pin}) and (\ref{Pout}) for $P_{\rm in}$ and $P_{\rm out}$, while $t_{\rm KL}$ and $e_{\rm max}$ are evaluated by numerical calculation except for Models IV6, VA6, VIA6, for which we use the formula (\ref{tKL}). The number $N$ defined by (\ref{N_accumulate}) denotes how many cycles the inner orbit evolves during highly eccentric stage.}
\label{model_and_observability}
\end{center}
\end{table}

\begin{figure}[h]
  \centering
  \includegraphics[width=8cm]{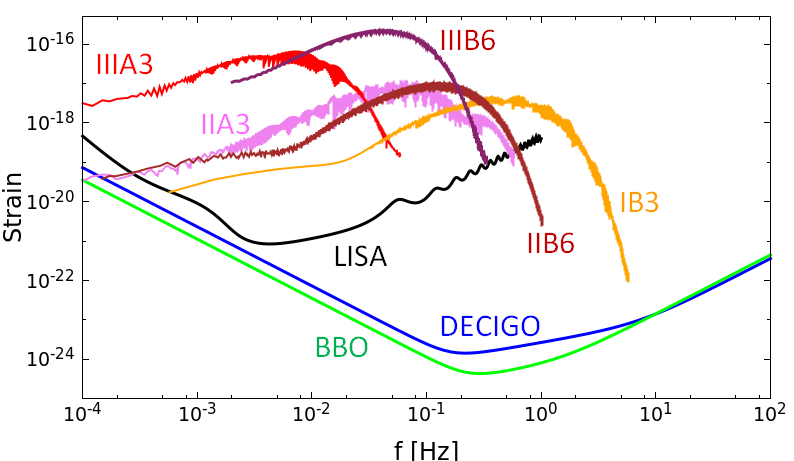}
  \hskip 1cm
  \includegraphics[width=8cm]{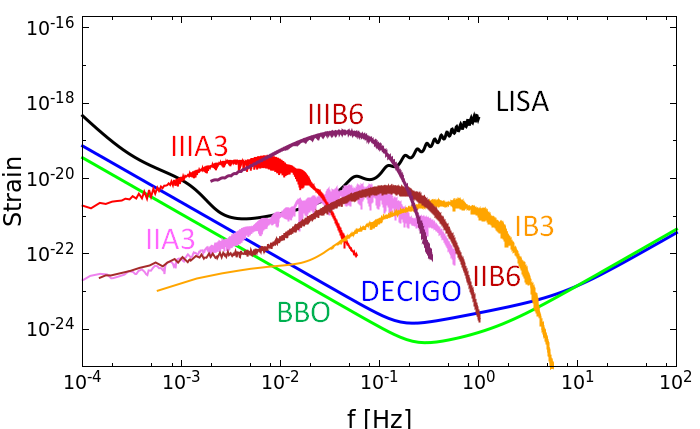}
\caption{\small The strain versus frequency plot for 
Models IB3, IIA3, IIB6, IIIA3 and IIIB6. The black, blue, and green  curves show  LISA, DECIGO, and BBO design sensitivity.  The observer distances 
are 10kpc (left) and 16 Mpc (right).}
\label{fig:hfplot_10kpc}
\end{figure}
\end{widetext}

We have shown the characteristic strain curves for  Models IA1, IA1$_3$ and IB6 in the text. Here we present the observability for the other models.
In Fig. \ref{fig:hfplot_10kpc},  we present 
the maximum strain curves for the above models 
(IB3, IIA3, IIB6, IIIA3, and IIIB6).
We find that all models can be observable 
if the GW sources are 
in our galaxy. On the other hand, when they are at the distance of 16 Mpc, although
Models IIIA3 and IIIB6 are still observable by LISA, 
the other models are observable only by DECIGO or BBO.

\end{document}